\documentclass[3p]{elsarticle}

\usepackage{xcolor,textcomp}
\usepackage{graphicx,hyperref,epstopdf,mathtools}
\usepackage{amsmath,amssymb,multirow,array,gensymb}
\usepackage{fancyhdr,float}
\usepackage[none]{hyphenat}
\usepackage{enumitem,relsize}
\usepackage{footmisc}
\usepackage{soul,subfigure,longtable}

\makeatletter
\def\ps@pprintTitle{%
 \let\@oddhead\@empty
 \let\@evenhead\@empty
 \def\@oddfoot{}%
 \let\@evenfoot\@oddfoot}
\makeatother

\begin{document}

\begin{frontmatter}


\title{A Quantitative Study of Energy Localization Characteristics in Defect-embedded Phononic Crystals}

\author[inst1]{Vinod Ramakrishnan\corref{cor1}}
\affiliation[inst1]{organization={The Grainger College of Engineering, Mechanical Science and Engineering},
  addressline={University of Illinois Urbana-Champaign},
  postcode={Illinois, 61801},
  city={Urbana},
  country={USA}}
\cortext[cor1]{Corresponding author}
\ead{vinodr@illinois.edu}
\author[inst1]{Kathryn H. Matlack}

\begin{abstract}

Phononic crystals (PnCs) are periodic engineered media that can customize the spatio-temporal characteristics of mechanical energy propagation. PnCs that additionally leverage precisely embedded defects can achieve robust energy localization with desirable spatio-temporal characteristics, opening avenues for critical engineering applications, e.g., energy harvesting, waveguiding, and fluid flow control. Numerous studies have qualitatively explored the localized dynamics via simulations and experiments, investigating the defect resonance frequency as the primary feature. However, the frequency represents only a subset of the relevant characteristics and a systematic approach to quantify the full scope of the defect dynamics remains elusive. This article establishes the frequency, mode shape, and localized velocity (or displacement) amplitude envelope as three significant factors governing the defect resonance dynamics, and quantitatively examines these characteristics using a modified version of the perturbed tridiagonal n-Toeplitz method. The proposed method accurately estimates the resonance characteristics in 1D and 2D defect-embedded PnC lattices with single and multiple defects and elucidates the effects of damping. The method is used to highlight how the key characteristics of defect modes depend on system parameters. Finally, we demonstrate the benefits of defect modes through two defect-based PnCs that can accommodate -- (i) a virtual ground, and (ii) achieve customized acoustic interaction and absorption, and use the proposed method to analyze these scenarios. The proposed strategy can be readily extended to more elaborate PnCs and augments the design space for defect-based PnCs.

\end{abstract}




\end{frontmatter}

\section{Introduction}\label{sec:Introduction}

Historically, the presence of defects or impurities in the crystalline lattice structures of natural materials has been considered an undesirable characteristic that can adversely impact the material properties. Consequently, the materials research community has developed numerous experimental techniques, e.g., X-ray diffraction\cite{WarrenCC1990}, ultrasound non-destructive evaluation\cite{DrinkwaterNDTEI2006,KimAM2021}, in the past century, to precisely identify the location, and study the impact of these structural (e.g., point\cite{LeibfriedSpringer2006}, line\cite{StonehamOXP2001}) defects within the material architecture on the meso- and macro-scale dynamics. The systematic progression in the study of defects has elucidated their critical role in determining material performance and gradually shifted the long-standing negative perception associated with them to a point where defects are now viewed as control parameters in materials design. In recent years, the idea of precisely engineering defects into a solid microstructure has gained tremendous traction forming the cornerstone of multiple modern fields of research, e.g., silicon-based chip manufacturing\cite{NishiCRC2000}, battery science\cite{NazriSSBM2008}, solid-oxide fuel cells\cite{HuangElsevier2009}.

Simultaneously, in the field of mechanics and materials, the advent of phononic crystals (PnCs) and metamaterials, primarily featuring periodic, cleverly architected mesoscale building blocks (i.e., unit cells) have offered an invaluable capability to mimic the ordered atomic architectures of natural materials at a more accessible scale and tailor the macroscale material behavior. PnCs and metamaterials have featured diverse unit cell designs, leveraging active elements\cite{WangSMS2017,AttarzadehPRAppl2020}, geometric instabilities\cite{BertoldiPRB2008,KochmannAMR2017}, multistability\cite{RamakrishnanJAP2020,ChenNature2021}, and kinematic amplification mechanisms\cite{FrazierJAP2022,RamakrishnanJASA2023}, eliciting peculiar material behavior, e.g., auxeticity\cite{AuricchioMD2019}, non-reciprocity\cite{AttarzadehPRAppl2020}, directional wave-guiding\cite{LiuAP2020}. Though this research is relatively recent, the community has shared identical views with the larger materials research community about the presence of structural or material defects\cite{RomeroNJP2010}. Similar to the prevalent perception of defects, this view has continuously evolved and, in recent years, has led to the exploration of material architectures with precisely embedded defects to robustly localize energy at desirable locations and spatio-temporal scales. Multiple strategies\cite{KainaSR2017,LiEML2016,LvCrys2019,MiniaciJAP2021,MeeussenNP2020} exist in the literature, to introduce defect modes in a variety of material architectures, e.g., Kaina \emph{et al.}\cite{KainaSR2017} experimentally demonstrated a waveguide allowing wave propagation at reduced group velocities by altering the lattice spacing of the periodic sub-wavelength resonant structures, Li \emph{et al.}\cite{LiEML2016} induced a defect mode in a soft bilayer by triggering a wrinkling instability at a prescribed location via strain-engineering, Xu-Feng \emph{et al.}\cite{LvCrys2019} experimentally demonstrated high energy localization and efficient energy harvesting by inducing geometric defects in a PnC beam. These proposals primarily present a qualitative analysis of the defect resonance dynamics via numerical simulations, parameter studies, supercell analysis, and experiments. 

Consequently, subsequent studies have explored analytical methods\cite{JoJAP2021,JoIJMS2022,JoMAMS2022,BastawrousJASA2022,HasanPRSA2019,HasanJEL2024,TargoffJAS1947,CaiJAM1995,CaiJSV2003,CaiJSV2005} to characterize the defect mode frequency in terms of the defect parameters, e.g., geometric parameters, material properties. Though these approaches have aided the development of more informed material design strategies, the frequency represents only a subset of the relevant dynamic characteristics of defect modes. Additional modal characteristics, e.g., mode shape, localized velocity (or displacement) amplitude envelope at the defect location, can also play a significant role in determining the effectiveness of defect-based PnCs and augment their functionality for wave-guiding\cite{ShelkeJIMSS2013}, energy harvesting\cite{AkbariSAA2023,ParkNE2019}, flow control\cite{HusseinPRSA2015} applications, motivating the need for a more comprehensive characterization of defect modes. Addressing this demand, we present a modified analytical approach to quantify three critical characteristics of defect modes, i.e., frequency, mode shape, and resonance velocity amplitude envelope, in a defect-embedded grounded monoatomic phononic crystal.

The remainder of the article is organized as follows -- Sec. \ref{sec:Model} provides a comprehensive analysis of the acoustic characteristics of the emergent bandgap resonances in a defect-embedded grounded monoatomic phononic crystal. Sec. \ref{sec:NumRes} presents two defect-embedded PnCs, demonstrating the benefits of defect modes. Finally, Sec. \ref{sec:Conclusion} presents a summary and the key takeaways from the article.

\section{Model Description}\label{sec:Model}

\begin{figure*}
    \centering
    \includegraphics{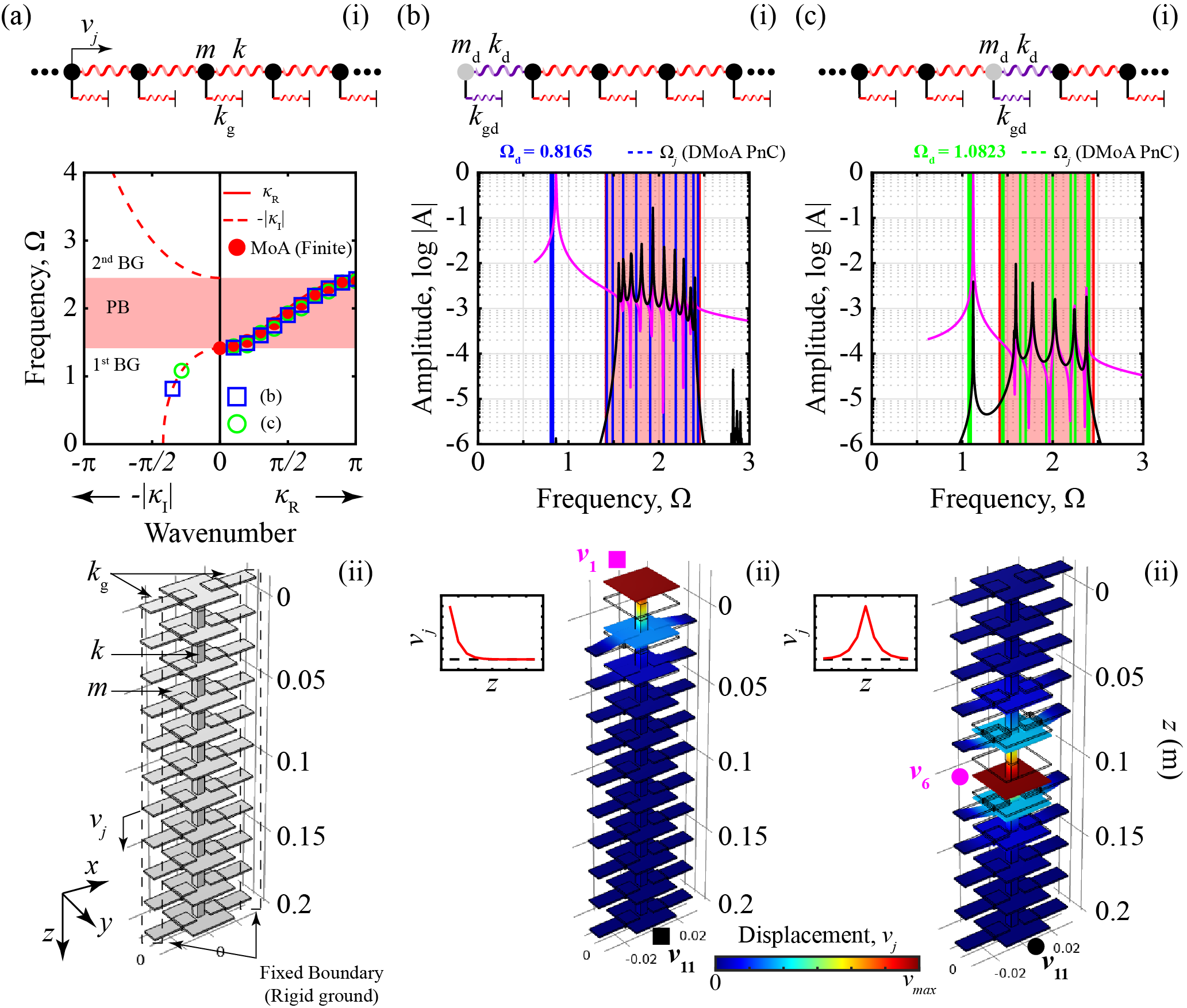}
    \caption{DMoA PnC. (a-i) A mass-spring MoA PnC model, dispersion curves with pass bands (PBs), band gaps (BGs), and eigenfrequencies of finite ($\mathrm{N}=11$) edge and bulk DMoA PnCs. (a-ii) A 3D structural model consisting of an array of aluminum plates ($m$), connected to their nearest neighbors using silicone beams ($k$) and individually grounded using Veroclear clips ($k_\mathrm{g}$) whose axial dynamics are represented by the MoA PnC model. (i) Mass-spring DMoA PnC models and frequency response functions (FRFs) obtained from 3D FEM studies of a finite defect-embedded 3D structure, (ii) defect eigenmode shapes featuring - (b) an edge defect, and (c) a bulk defect.}
    \label{fig:FIG1}
\end{figure*}

Consider a composite 3D structure shown in Fig. \ref{fig:FIG1}(a-ii). The axial dynamics of the 3D structure can be approximated as a periodic lumped mass-spring grounded monoatomic phononic crystal (MoA PnC) model (Fig. \ref{fig:FIG1}(a-i)) with mass, $m$, interaction stiffness, $k$, and onsite grounding stiffness, $k_\mathrm{g}$ (see Sec. \ref{sec:App_COMSOL}). The non-dimensional equation of motion for the $j^\mathrm{th}$ mass in an infinite MoA PnC is given by,
\begin{equation}
    v_{j,\bar{t}\bar{t}}+\left(2v_j-v_{j+1}-v_{j-1}\right)+k_\mathrm{g,r}v_j=0,
    \label{eq:MoA_EOM}
\end{equation}
where, $v_j$ is the axial displacement of the $j^\mathrm{th}$ mass, $\bar{t}=\omega_0 t$ is the non-dimensional time, $\omega_0=\sqrt{k/m}$ is the reference temporal frequency and $k_\mathrm{g,r}=k_\mathrm{g}/k$ is the grounding stiffness ratio. Consequently, the dispersion relation for the MoA PnC, assuming a plane wave \emph{ansatz} of the form $v_j=\tilde{v}_j e^{i(\kappa x_j-\Omega \bar{t})} \ \forall \ j=1,2,\cdots,\mathrm{N}\ \mathrm{and} \ x_j=j-1$, is calculated as\cite{NarisettiJVA2010,NadkarniPRE2014}:
\begin{equation}
    \Omega(\kappa)=\sqrt{k_\mathrm{g,r}+4\sin^2\left(\frac{\kappa}{2}\right)},
    \label{eq:MoA_Dispersion}
\end{equation}
where, $\Omega$ is the non-dimensional temporal frequency, and $\kappa=\kappa_R+i\kappa_I \ \forall \ \kappa_R,\kappa_I \in \mathbb{R}$, is the non-dimensional complex wavenumber. The dispersion curves (Eq. \eqref{eq:MoA_Dispersion}, $k_\mathrm{g,r}=2$) and the natural eigenfrequencies, $\Omega_j$ of a finite MoA PnC with $\mathrm{N}$ degrees of freedom and free boundaries are plotted in Fig. \ref{fig:FIG1}a, highlighting the band gap (BG, $\kappa_R=0 \ \mathrm{or} \ \pi, \kappa_I\neq0$) and the pass band (PB, $\kappa_R\neq0, \kappa_I=0$) frequency ranges. The grounding elastic springs are the primary source of the bounded BG between $\Omega=0$ and a finite $\Omega=\sqrt{k_\mathrm{g,r}}$. The grounding springs resist the deformation of masses with respect to a fixed ground and participate in all the eigenmodes describing the MoA PnC dynamics. Consequently, the PB range and specifically the lower edge of the PB, $\Omega(\kappa=0)=\sqrt{k_\mathrm{g,r}}$ is directly linked to the elasticity of these grounding elements, opening up a BG below this modal frequency.

When the periodicity of a finite MoA PnCs is altered by introducing a change in one or more material properties (defect), i.e., $m_\mathrm{r}=m_\mathrm{d}/m$, $k_\mathrm{r}=k_\mathrm{d}/k$, or $k_\mathrm{gd,r}=k_\mathrm{gd}/k$ at a specified location within the MoA PnC, the eigenfrequency (or multiple eigenfrequencies, see Sec. \ref{sec:App_MultiBGRes}) associated with the lower (upper) edge of the PB migrates to the lower (upper) BG, creating a BG resonance at $\Omega_1=\Omega_\mathrm{d}<\sqrt{k_\mathrm{g,r}}$ ($\Omega_\mathrm{N}=\Omega_\mathrm{d}>\sqrt{4+k_\mathrm{g,r}}$). Note that a BG resonance can exist even in an ungrounded mass-spring PnCs\cite{HasanJEL2024,BastawrousJASA2022,MachadoAIAASci2024} however, a grounded PnC ($k_\mathrm{g,r}\neq0$) provides an opportunity to engineer a BG resonance in the lower BG, $\Omega_\mathrm{d}\in[0,\sqrt{k_\mathrm{g,r}})$, that manifests as the first eigenmode of the altered MoA PnC. The existence of a BG resonance as the principal, low-frequency eigenmode may be beneficial in ensuring their effective participation in the system dynamics, facilitating defect-based PnCs for practical applications that can involve complex operational requirements, e.g., response to low-frequency broadband external forces\cite{MachadoAIAASci2024}, amplitude-phase requirements\cite{HusseinPRSA2015}. We refer to the altered MoA PnC as a defect-embedded grounded monoatomic phononic crystal (DMoA PnC) and the resultant BG resonance as a defect mode for brevity.

The DMoA PnCs can be explored in two fundamentally distinct configurations --  (i) an edge DMoA PnC where material properties at one of the boundaries of the finite MoA PnC are altered and, (ii) a bulk DMoA PnC\cite{CaiJAM1995} where material properties in the bulk deviate from the periodic properties of the MoA PnC. The natural frequencies of these DMoA PnCs are obtained by solving the matrix eigenvalue problem, $\mathbf{K}_\mathrm{d}\hat{\mathbf{V}}=\mathbf{0}$ (see Sec. \ref{sec:App_EdgeDefect}), with $m_\mathrm{r}=k_\mathrm{r}=1$ and $k_\mathrm{gd,r}=0$ and, overlaid onto the dispersion plot in Fig. \ref{fig:FIG1}a, indicating the presence of a defect mode (BG resonance) in both cases albeit, with distinct frequencies. 

The frequency response functions (FRFs) of the 3D structures with edge and bulk defects, plotted in Fig. \ref{fig:FIG1}(b,c-i) are calculated by applying a uniform boundary load in the axial (z-) direction to the defect plate and performing a 3D FEM frequency analysis in COMSOL. Though modes of other polarizations (beam bending modes, see Fig. \ref{fig:SFIG5}) are faintly excited in the edge defect-embedded structure ($\Omega \approx 2.9$), the FRF resonance peaks primarily denote the axial resonance modes of the structures, showing an excellent correlation with the eigenfrequencies calculated from the edge and bulk DMoA PnC mass-spring models. In addition, the FRF captures all $\mathrm{N}$ axial eigenmodes and only $(\mathrm{N}+1)/2$ axial eigenmodes, respectively, in the edge and bulk defect cases as the excitation location is at the structural boundary in the former as compared to the material bulk in the latter.

The 3D structures manifest higher FRF amplitudes at the defect resonance frequency, $\Omega_\mathrm{d}$ compared to other PB eigenfrequencies. A reduction in amplitudes between the magenta ($|\mathrm{A}|_{j=d}$) and black curves ($|\mathrm{A}|_{j=\mathrm{N}}$) in Fig. \ref{fig:FIG1}(b,c-i) at $\Omega=\Omega_\mathrm{d}$ also indicates a strong localization of energy at the defect owing to an exponentially decaying eigenmode shape away from the defect location (see Fig. \ref{fig:FIG1}(b,c-ii)). Since the axial modes of the 3D structures subjected to uniform axial loads are accurately captured by the reduced order model, the remainder of the article analytically and numerically investigates the DMoA PnC mass-spring model with material parameters -- $k_\mathrm{g,r}=2$, $m_\mathrm{r}=1$, $k_\mathrm{r}=1$ and $k_\mathrm{gd,r}=0$ unless explicitly stated otherwise.

The defect frequency, $\Omega_\mathrm{d}$, and the exponential decay rate, $\kappa_I$ comprehensively describe the contribution of the material defect in altering the base MoA PnC dynamics. In addition, although rarely considered in prior studies, we quantify and describe a third critical parameter -- the velocity amplitude envelope at resonance, $\dot{\mathrm{V}}_\mathrm{E}$ that can promote defect-based PnCs for novel applications, e.g., energy harvesting\cite{AkbariSAA2023,ParkNE2019}, fluid flow control\cite{HusseinPRSA2015,WilleyJFS2023}, where response amplitude of the PnC can play a major role in determining operational efficiency. In the following sections, we elaborate on a modified analytical approach to quantify these critical characteristics of the defect dynamics of the DMoA PnC -- $\{\Omega_\mathrm{d},\kappa_I,\dot{\mathrm{V}}_\mathrm{E}\}$. The proposed approach can be readily extended to general boundary conditions\cite{BastawrousJASA2022,HasanJEL2024} and mass-spring PnC models comprising unit cells with more degrees of freedom, e.g., defect-embedded diatomic\cite{JoIJMS2022,BastawrousJASA2022} and polyatomic\cite{HasanPRSA2019} phononic crystals.

\subsection{Defect eigenfrequency and eigenmode shape}\label{sec:DMoA}

\begin{figure*}
    \centering
    \includegraphics{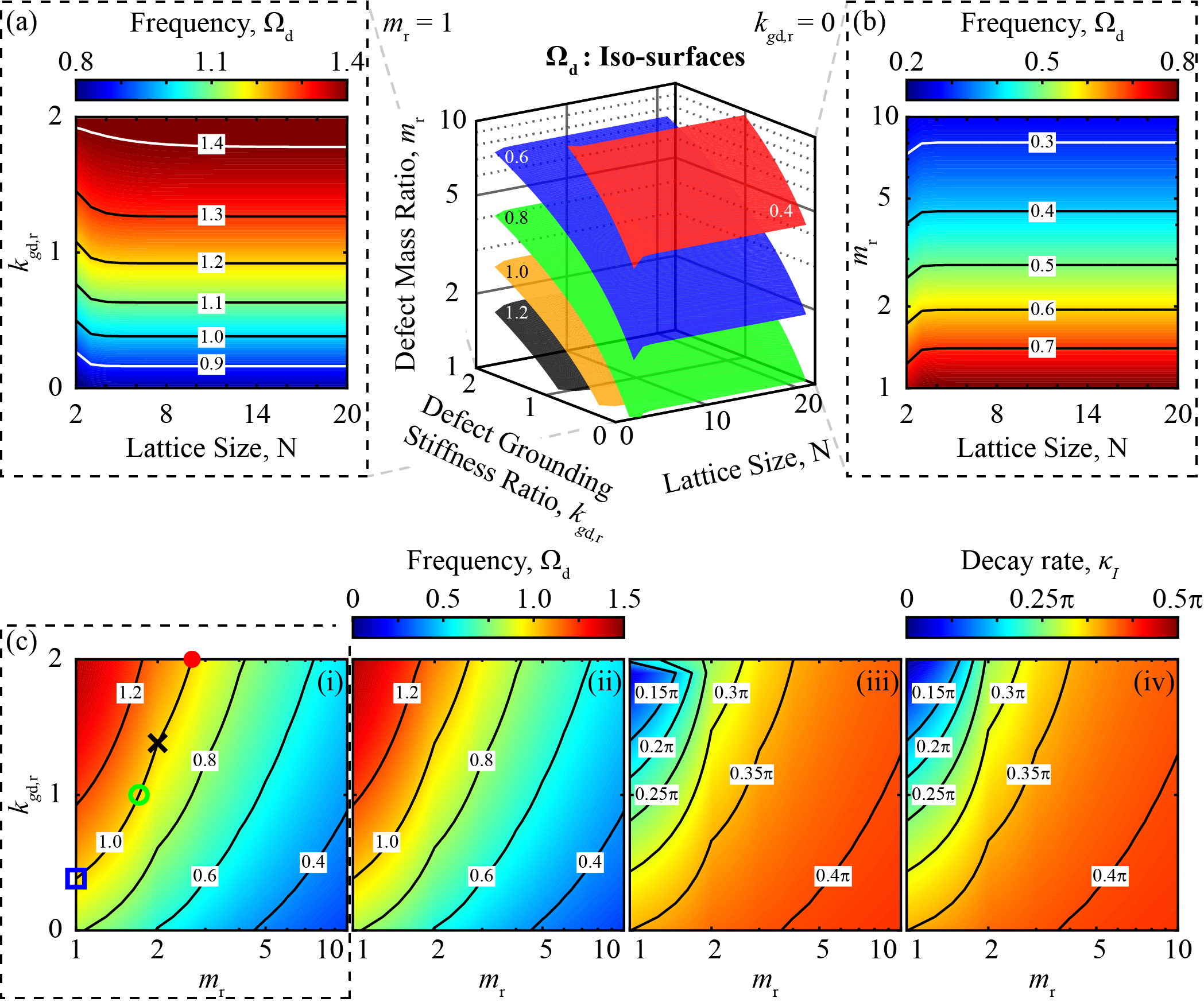}
    \caption{Iso-curves of the defect eigenmode characteristics. (Center) 3D plot of the iso-surfaces of defect frequency, $\Omega_\mathrm{d}=\{0.4,0.6,0.8,1.0,1.2\}$ as a function of material properties, $m_\mathrm{r}$ (log-scale) and $k_\mathrm{gd,r}$ and, the size of the base MoA PnC, $\mathrm{N}$. (a) Contour depicting a 2D slice of the 3D central figure, taken at $m_\mathrm{r}=1$, (b) Contour depicting a 2D slice of the 3D central figure, taken at $k_\mathrm{gd,r}=0$, (c-i) Contour depicting a 2D slice of the 3D central figure, taken at $\mathrm{N}=21$, (c-ii) Contour plotting the analytical solution for defect frequency, $\Omega_\mathrm{d}$ as a function of $\{m_\mathrm{r},k_\mathrm{gd,r}\}$, (c-iii) Numerical and (c-iv) analytical contours of the defect eigen mode shape, i.e., decay rate, $\kappa_I$ as a function of $\{m_\mathrm{r},k_\mathrm{gd,r}\}$. The $\{\square,\mathbf{O},\mathbf{\times},\bullet\}$-markers indicate DMoA PnC configurations investigated in Fig. \ref{fig:FIG3}.}
    \label{fig:FIG2}
\end{figure*}

Consider the edge DMoA PnC model illustrated in Fig. \ref{fig:FIG1}b. The edge defect resembles a local resonator appended to the boundary of a finite, homogeneous elastic foundation (base MoA PnC). The presence of the edge defect creates a localized defect mode in the altered material configuration that can be modeled as a reduced order edge DMoA PnC illustrated in Fig. \ref{fig:FIG1}b. The BG eigenfrequencies of a finite edge DMoA PnC are numerically calculated and plotted as iso-defect frequency surfaces, i.e., $\Omega_\mathrm{d}=\mathrm{const.}$, in the central 3D-plot in Fig. \ref{fig:FIG2} for varying degrees of freedom, $\mathrm{N}$, and material properties at the edge defect, i.e., $k_\mathrm{gd,r}$ and $m_\mathrm{r}$, to understand the qualitative dependence of the defect frequency, $\Omega_\mathrm{d}$ on these parameters. Fig. \ref{fig:FIG2}a presents a 2D slice of the central plot taken at $m_\mathrm{r}=1$. Consistent with the qualitative relation between frequency and stiffness, the magnitudes of the iso-defect frequency curves on the contour plot indicate that, $\Omega_\mathrm{d}$ increases as $k_\mathrm{gd,r}$ increases from $0\longrightarrow k_\mathrm{g,r}$. Similarly, in Fig. \ref{fig:FIG2}b (2D slice taken at $k_\mathrm{gd,r}=0$) the magnitude of iso-defect frequency curves decrease with increasing $m_\mathrm{r}$ ($>1$), as anticipated qualitatively. In addition, we notice that the iso-defect frequency curves in both Fig. \ref{fig:FIG2}a,b tend to asymptote to a constant value of $k_\mathrm{gd,r}$ and $m_\mathrm{r}$, respectively, at large $\mathrm{N}$. This observation provides valuable insight for our analytical model, where we can arbitrarily choose a large DMoA PnC size ($\mathrm{N}$) to obtain an accurate prediction for $\Omega_\mathrm{d}$, independent of the actual size of the finite DMoA PnC. 

Finally, Fig. \ref{fig:FIG2}(c-i,iii) plots the $\{\Omega_\mathrm{d},\kappa_{I}\}$, calculated from matrix eigen-analysis of a finite DMoA PnC ($\mathrm{N}=21$) as a function of $\{m_\mathrm{r},k_\mathrm{gd,r}\}$. A closer look at the matrix eigenvalue problem reveals a tridiagonal matrix form for the dynamic stiffness, $\mathbf{K}_\mathrm{d}$ (Sec. \ref{sec:App_EdgeDefect}) that is consistent with the general form of a perturbed tridiagonal Toeplitz matrix\cite{FonsecaAMS2007,BastawrousJASA2022}. The characteristic equation for $\mathbf{K}_\mathrm{d}$ can be written in the implicit form:

\begin{equation}
    f(\Omega)=g(\Omega),
    \label{eq:DMoA_MCE}
\end{equation}
where, without loss of generality for $\mathrm{N}=2n+1$ and $n\in\mathbb{N}$, 
\begin{equation*}
    f(\Omega)=\frac{\sin\left([n+1]\left[\cos^{-1}\left(\frac{d^2}{2a^2}-1\right)\right]\right)}{\sin\left(n\left[\cos^{-1}\left(\frac{d^2}{2a^2}-1\right)\right]\right)},
\end{equation*}
\begin{equation*}
    g(\Omega)=-\frac{\alpha\beta d + (\alpha+\beta)a^2}{(d+\alpha+\beta)a^2},
\end{equation*}
and, 
\begin{equation*}
    d=2+k_\mathrm{g,r}-\Omega^2,a=-1, \beta=-1,
\end{equation*}
\begin{equation*}
    \alpha=(k_\mathrm{gd,r}-k_\mathrm{g,r})-1+(1-m_\mathrm{r})\Omega^2,
\end{equation*}
represent the matrix entries of $\mathbf{K}_\mathrm{d}$. Eq. \eqref{eq:DMoA_MCE} can be solved numerically, assuming a large value of $n$ (i.e, large $\mathrm{N}$) to estimate $\{\Omega_\mathrm{d},\kappa_I\}$. Fig. \ref{fig:FIG2}(c-ii,iv) plot the $\{\Omega_\mathrm{d},\kappa_I\}$ contours, respectively, obtained using the analytical approach, showing an excellent correlation with their numerical counterparts in Fig. \ref{fig:FIG2}(c-i,iii). Note that the decay rate, $\kappa_I$ only depends on the base MoA PnC properties $k_\mathrm{g,r}$ and defect frequency, $\Omega_\mathrm{d}$. Therefore, each $\Omega_\mathrm{d}$ is associated with a unique $\kappa_I$ leading to identical shapes for iso-defect frequency and iso-decay rate curves in Fig. \ref{fig:FIG2}c.

Alternately, for the bulk DMoA PnC (Fig.\ref{fig:FIG1}c), we utilize the U-transformation method\cite{CheungJEM1989} proposed by Cai \emph{et al.}\cite{CaiJAM1995} to arrive at an implicit equation identical to Eq. \eqref{eq:DMoA_MCE} that can be solved numerically to estimate the defect frequency and mode shape. Though an explicit analytical form of $\{\Omega_\mathrm{d},\kappa_I\}$ remains elusive, the proposed solution approach for both DMoA PnCs require a numerical calculation only at the terminal step to solve a well-defined implicit analytical equation (Eq. \eqref{eq:DMoA_MCE}). Therefore, the analytical approach executed with a numerical algorithm\cite{LagariasSIAMJO1998,MiettinenKAP1999} of our choice allows a systematic derivation of DMoA PnC material parameters that produce the desired spatio-temporal defect characteristics.

\subsection{Velocity amplitude envelope at defect mode resonance}\label{sec:DMoA Velocity}

\begin{figure*}
    \centering
    \includegraphics{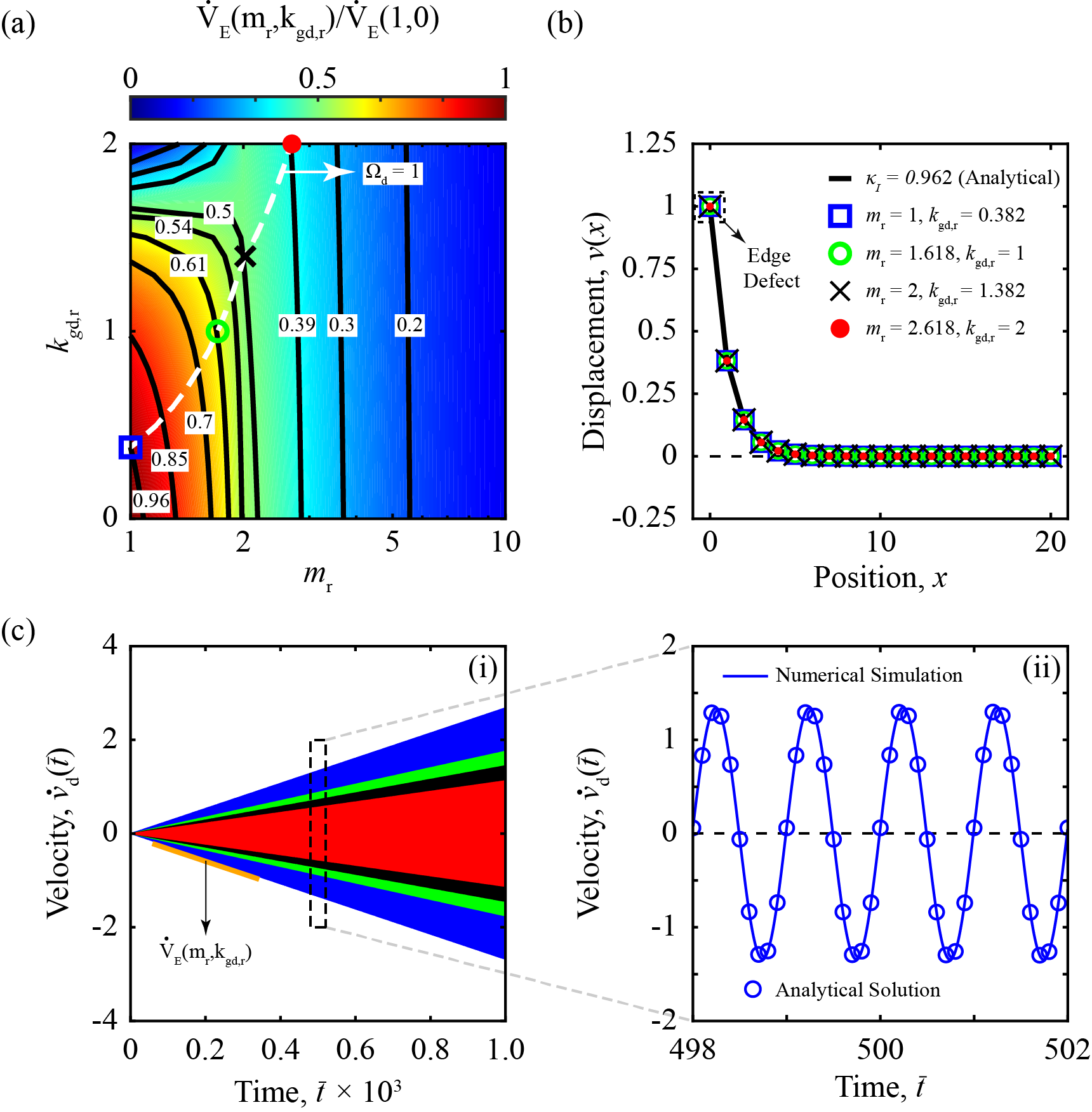}
    \caption{Defect Mode shape and response of edge defect at resonance. (a) Normalized velocity amplitude envelope, $\dot{\mathrm{V}}_\mathrm{E}(m_\mathrm{r},k_\mathrm{gd,r})/\dot{\mathrm{V}}_\mathrm{E}(1,0)$ contour plot with iso-velocity amplitude envelope curves and the iso-defect frequency curve at $\Omega_\mathrm{d}=1$ from Fig. \ref{fig:FIG2}(c-i).(b) Defect mode shapes for the $\{m_\mathrm{r},k_\mathrm{gd,r}\}$ pairs indicated by the respective markers, lying on the $\Omega_\mathrm{d}=1$ and $\kappa_I=0.962$ in Fig. \ref{fig:FIG2}c, normalized by the edge defect displacement, $v_1(\bar{t}) \equiv v_\mathrm{d}(\bar{t})$ , (c-i) Velocity response of the edge defect, $\dot{v}_\mathrm{d}(\bar{t})$ to a resonant force, $\mathrm{F}(\bar{t})=\mathrm{F}_0\sin (\Omega_\mathrm{d}\bar{t})$, (c-ii) Magnified view of the numerical and analytical velocity response (Eq. \eqref{eq:ResVelocity}) for $\{m_\mathrm{r},k_\mathrm{gd,r}\}=\{1,0.382\}$.}
    \label{fig:FIG3}
\end{figure*}

A closer look at the iso-defect frequency and iso-decay rate curves in Fig. \ref{fig:FIG2}c indicates the presence of multiple $\{m_\mathrm{r},k_\mathrm{gd,r}\}$ pairs that can manifest the same $\{\Omega_\mathrm{d},\kappa_I\}$. Therefore, we formulate and study an additional quantity associated with resonance dynamics at the defect location, i.e., the velocity amplitude envelope, $\dot{\mathrm{V}}_\mathrm{E}$ (see Sec. \ref{sec:RVAE}) to distinguish the dynamic behavior of edge DMoA PnCs with distinct material properties, $\{m_\mathrm{r},k_\mathrm{gd,r}\}$. Absolute velocity (displacement) amplitudes at resonance have rarely been featured in fundamental research studies of PnCs. However, their emerging relevance in determining PnC efficiency for applications, e.g., energy harvesting\cite{AkbariSAA2023,ParkNE2019}, flow control\cite{HusseinPRSA2015,WilleyJFS2023} motivates their consideration as a critical behavioral parameter describing the PnC dynamics. Hence, a prescribed $\dot{\mathrm{V}}_\mathrm{E}$ is considered in this section as the third and final parameter to complete the edge DMoA PnC design.

Fig. \ref{fig:FIG3}a plots the normalized velocity amplitude envelope (Eq. \eqref{eq:Resonance_Coeff1}) as a function of $\{m_\mathrm{r},k_\mathrm{gd,r}\}$. The iso-defect frequency curve at $\Omega_\mathrm{d}=1$ intersects the iso-velocity amplitude envelope curves at a single point in the given range of $\{m_\mathrm{r},k_\mathrm{gd,r}\}$. Consequently, optimizing for a desired $\{\Omega_\mathrm{d}, \kappa_I,\dot{\mathrm{V}}_\mathrm{E}\}$ is a well-posed problem that will produce a unique solution, e.g., the $\{\square,\mathbf{O},\mathbf{\times},\bullet\}$-markers indicating $\{m_\mathrm{r},k_\mathrm{gd,r}\}$ values corresponding to $\Omega_\mathrm{d}=1$, $\kappa_I=0.962$, and $\dot{\mathrm{V}}_\mathrm{E}/\dot{\mathrm{V}}_\mathrm{E}(1,0)=\{0.96,0.61,0.5,0.39\}$, respectively.

As anticipated, the mode shapes associated with DMoA PnCs described by the $\{\square,\mathbf{O},\mathbf{\times},\bullet\}$-markers, plotted in Fig. \ref{fig:FIG3}b manifest an identical decay rate, $\kappa_I=0.962$. Each DMoA PnC is now subject to a sinusoidal external force, $\mathrm{F}(\bar{t})=\mathrm{F}_0\sin (\Omega_\mathrm{d}\bar{t})$ at the edge defect mass, and Fig. \ref{fig:FIG3}(c-i) plots the corresponding numerical and analytical velocity response, 
\begin{equation}
    \dot{v}_\mathrm{d}(\bar{t})=\dot{\mathrm{V}}_\mathrm{E}\bar{t}\sin\left(\Omega_\mathrm{d}\bar{t}\right),
    \label{eq:ResVelocity}
\end{equation}
for a total simulation time is $\bar{t}_\mathrm{sim}=1000$.

Consistent with a typical response at resonance, all DMoA PnCs exhibit an oscillatory velocity response at the prescribed defect frequency, $\Omega_\mathrm{d}=1$ with a linearly increasing velocity amplitude in time. However, each DMoA PnC manifests a unique rate of increase in velocity amplitude that we designate as the velocity amplitude envelope, $\dot{\mathrm{V}}_\mathrm{E}$. $\dot{\mathrm{V}}_\mathrm{E}$ can be analytically estimated in terms of the $\{\Omega_\mathrm{d}, \kappa_I\}$ or $\{m_\mathrm{r}, k_\mathrm{gd,r}\}$, and the input force amplitude, $\mathrm{F}_0$ (see Sec.\ref{sec:RVAE}). Fig. \ref{fig:FIG3}(c-ii) contrasts the analytical resonance solution in Eq. \eqref{eq:ResVelocity} and the velocity response recorded in the numerical simulations ($\square$ - marker, $\{m_\mathrm{r},k_\mathrm{gd,r}\}=\{1,0.382\}$), showing an excellent correlation between the two. Therefore, the proposed analytical approach offers a broad set of design parameters to construct a customized DMoA PnC based on operational requirements.

\subsection{Damped DMoA PnCs}
The analytical approach described above can be extended to DMoA PnCs containing dissipative elements, by taking into account the damping matrix, $\mathbf{C}$ and modifying the dynamic stiffness matrix, $\mathbf{K}_\mathrm{d}=\mathbf{K}+i\Omega\mathbf{C} -\Omega^2\mathbf{M}$, resulting in matrix coefficients:  
    \begin{equation*}
        d=2+k_\mathrm{g,r}+i\Omega(c_\mathrm{on,r}+2c_\mathrm{in,r})-\Omega^2,a=-(1+i\Omega c_\mathrm{in,r}), \beta=-(1+i\Omega c_\mathrm{in,r}),
    \end{equation*}
    \begin{equation*}
        \alpha=(k_\mathrm{gd,r}-k_\mathrm{g,r})-(1+i\Omega c_\mathrm{in,r})+(1-m_\mathrm{r})\Omega^2;
    \end{equation*}
where, the general damping matrix, $\mathbf{C}=c_\mathrm{on,r}\mathbf{I}+c_\mathrm{in,r}\mathbf{K}_\mathrm{in}$ (see Sec. \ref{sec:App_EdgeDefect}), 
accommodates both onsite and intersite damping elements - $c_\mathrm{on,r}=c_\mathrm{on}/\sqrt{km}$ and $c_\mathrm{in,r}=c_\mathrm{in}/\sqrt{km}$ , respectively. Substituting the above coefficients into, Eq. \eqref{eq:DMoA_MCE} will provide the defect mode frequencies and mode shapes in the damped DMoA PnCs.

In addition, in the presence of damping ($c_\mathrm{on,r}\neq0$ and/or $c_\mathrm{in,r}\neq0$), the DMoA PnC produces a bounded steady-state response, 
    \begin{equation*}
    \dot{v}_\mathrm{d}=\dot{\mathrm{V}}_\mathrm{SS}\left[\alpha\Omega_\mathrm{d}\sin(\Omega_\mathrm{d}\bar{t})+\cos(\Omega_\mathrm{d}\bar{t})\right],
    \end{equation*}
at the edge defect, when subject to external forcing, $\mathrm{F}(\bar{t})=\mathrm{F}_0\sin (\Omega_\mathrm{d}\bar{t})$. Therefore, the steady-state velocity amplitude, $\dot{\mathrm{V}}_\mathrm{SS}=-b_I\Omega_\mathrm{d}\widetilde{\mathrm{V}}_\mathrm{d}$ (Sec. \ref{sec:App_SS_VelAmp}) can be utilized as an alternate defect mode characteristic to the resonance velocity amplitude envelope, $\dot{\mathrm{V}}_\mathrm{E}$ in damped DMoA PnCs.

\subsection{Multi-defect embedded grounded monoatomic phononic crystal}\label{sec:Multi-DMoA}

\begin{figure*}
    \centering
    \includegraphics{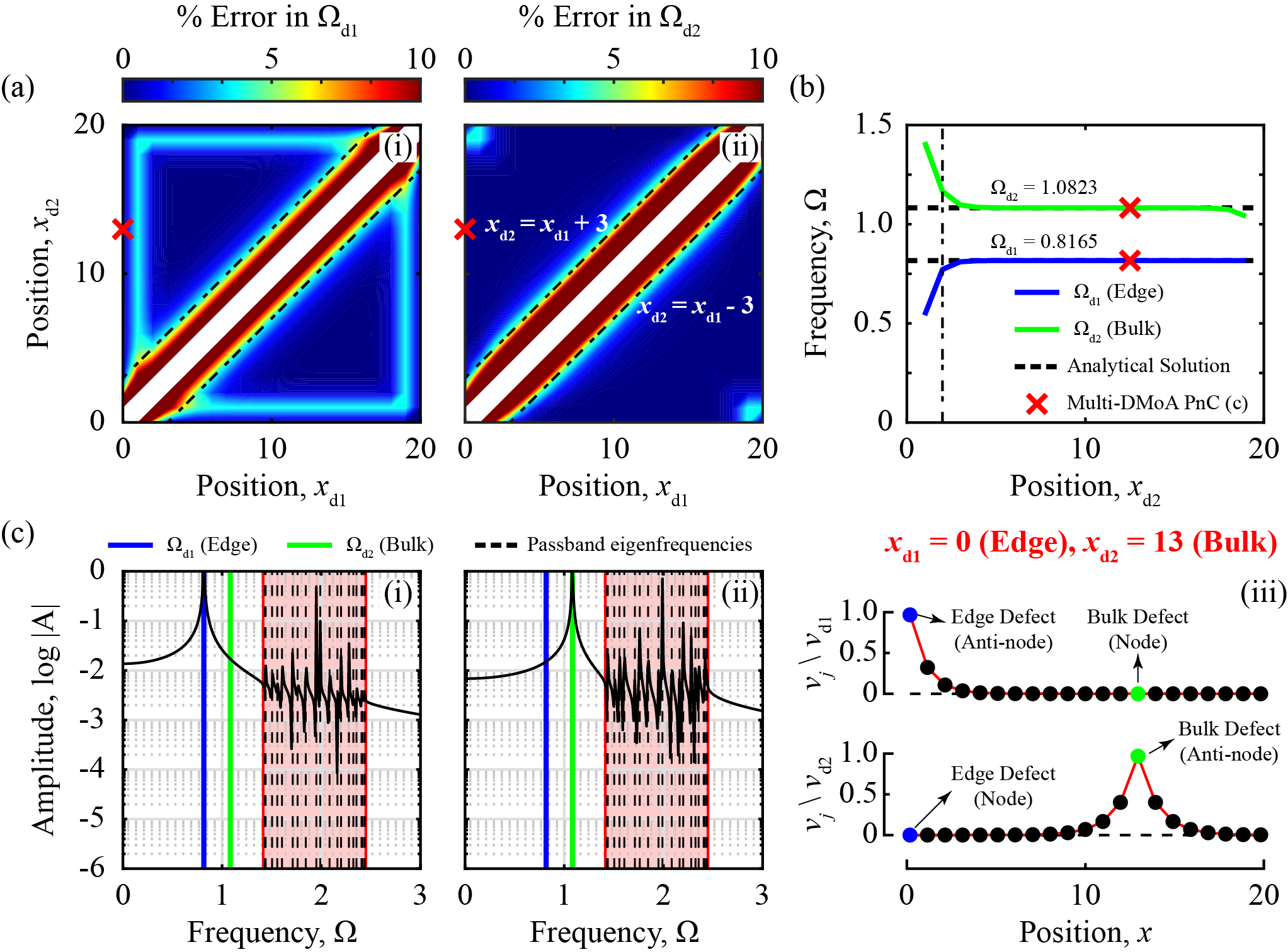}
    \caption{Multi-DMoA PnCs. (a) Absolute Percentage error contours for a Multi-DMoA PnC ($\mathrm{N}=21$) with two defects at $x_\mathrm{d1}, x_\mathrm{d2} \in [0,\mathrm{N}-1]$ and $x_\mathrm{d1} \neq x_\mathrm{d2}$. (b) Edge and bulk defect frequencies, $\Omega_\mathrm{d1}$ and $\Omega_\mathrm{d2}$, respectively, of a Multi-DMoA PnC with an edge defect at $x_\mathrm{d1}=0$ and a bulk defect at $x_\mathrm{d2}\in[1,\mathrm{N}-2]$. FRFs of a Multi-DMoA PnC denoted by \scalebox{1.25}{\bfseries $\times$}-marker in (a,b) when the (c-i) edge and (c-ii) bulk defect masses are excited and, their displacement response monitored, (c-iii) eigenmode shapes of the Multi-DMoA PnC.}
    \label{fig:FIG4}
\end{figure*}

In Sec. \ref{sec:DMoA} and Sec. \ref{sec:DMoA Velocity} we have established the accuracy of the proposed analytical approach in determining the dynamic characteristics of a DMoA PnC with a single defect. But, what are the limitations of the analytical approach? Can the approach be used to characterize DMoA PnCs with multiple defects? To address these questions and investigate the robustness of the solution procedure, in this section, we contrast the numerical and analytical dynamic characteristics of a multi-defect embedded grounded monoatomic phononic crystal (Multi-DMoA PnC).

Fig. \ref{fig:FIG4}a plots the absolute percentage error between the analytical and numerical values of the two defect frequencies, $\Omega_\mathrm{d1}$ and $\Omega_\mathrm{d2} (>\Omega_\mathrm{d1})$ generated due to the presence of two defects at locations, $x_\mathrm{d1}$ and $x_\mathrm{d2}$ in a Multi-DMoA PnC ($\mathrm{N}=21$). The diagonal direction in both these plots remains unpopulated as we assume distinct locations for the embedded material defects, i.e., $x_\mathrm{d1}\neq x_\mathrm{d2}$. The sea of dark blue, low error values ($0.0005\%-2\%$) in both plots indicate the validity of the analytical approach derived in Sec. \ref{sec:DMoA} for a large section of the Multi-DMoA PnC configurations provided, the two defects in the Multi-DMoA PnC remain relatively isolated from each other. Alternately, as the defect locations become proximal, i.e., within 3 masses, marked by the off-diagonal lines, $x_\mathrm{d2}=x_\mathrm{d1}\pm3$, the Multi-DMoA PnC configuration deviates from the isolated (single) defect assumption giving rise to a region of high error ($> 2\%$, max. error $\sim33\%$). However, note that the expanse of the high error region is dependent on the decay rate, $\kappa_I$ associated with bulk defect mode, i.e., as the $\kappa_I$ grows larger, the high error region shrinks (Sec. \ref{sec:App_MultiDef}). Therefore, an alternate approach is required to investigate the design space of proximal defects. This lies outside the scope of the current study. Although, prior efforts have studied PnCs with proximal defects via super-cell analysis\cite{JoIJMS2022,DengCrys2019,HuPB2014,Deymier2013}, that can provide good qualitative insights into a potential solution approach. Since we are primarily interested in localized BG resonances, analyzing an appropriately sized super-cell around the proximal defects, will reveal defect modes whose frequency and shape correlate to the complete PnC counterpart, as masses far away from the defect location remain (virtually) undeformed at defect resonance.

Another important observation to highlight in the $\Omega_\mathrm{d1}$ error plots in Fig. \ref{fig:FIG4}a, is the linear light blue regions - $x_\mathrm{d1}=1,\mathrm{N}-2$ and $x_\mathrm{d2}=1,\mathrm{N}-2$. These lines correspond to Multi-DMoA PnC configurations where one of the penultimate masses is a defect. Technically this defect can be categorized as a bulk defect, however, the proximity to the finite boundary of the Multi-DMoA PnC has a significant impact on the eigenmode characteristics, leading to a slightly higher error ($\sim 3.9\%$) compared to the Multi-DMoA PnC configurations in the dark blue regions. The impact of boundary effects is less prevalent in cases where the defect frequencies lie in the $2^\mathrm{nd}$ BG of the MoA PnC (see Sec. \ref{sec:App_MultiDef}). We can attribute this phenomenon to the real part of the wavenumber, $\kappa_R$ of these defect modes. The defect mode in the $1^\mathrm{st}$ BG has $\kappa_R=0$ that requires the boundary adjacent to the defect location to vibrate in-phase with the defect at resonance, leading to a rigid body-like motion of the boundary and defect masses. Alternately, the $2^\mathrm{nd}$ BG requires $\kappa_R=\pi$, thus, allowing the boundary mass to vibrate out of phase with the defect leading to relatively decoupled, anti-symmetric dynamics between the sites rendering the actual defect frequency for the latter case closer to the analytical estimate than the former.

\begin{figure*}
    \includegraphics{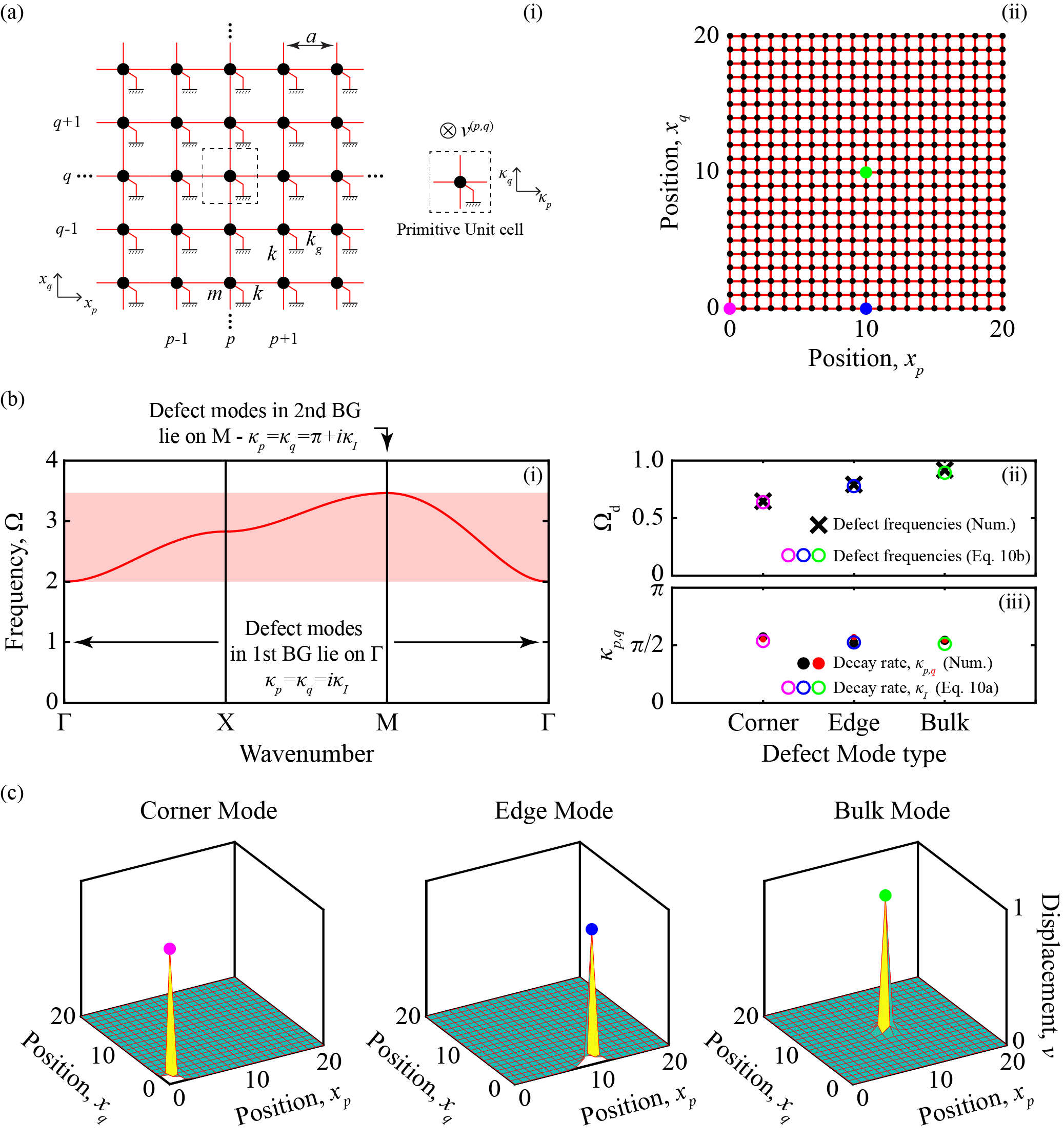}
    \caption{2D Multi-DMoA PnC. (a-i) 2D MoA PnC square lattice, the primitive unit cell, and an equivalent axis-symmetric configuration of a mass at $(x_p,x_q)$. (a-ii) A finite 2D DMoA PnC square lattice ($\mathrm{N}_p=\mathrm{N}_q=21$) with a corner (magenta), edge (blue), and a bulk (green) defect. (b-i) 2D dispersion curve (Eq. \eqref{eq:2D_Dispersion}) traversing the irreducible Brillouin zone $\Gamma-\mathrm{X}-\mathrm{M}-\Gamma$, i.e., $\kappa_p=0\rightarrow\pi,\kappa_q=0;\ \kappa_q=0\rightarrow\pi,\kappa_p=\pi;\ \kappa_p=\kappa_q=\pi\rightarrow0$. (b-ii) Defect frequency and (b-iii) decay rate predicted by Eq. \eqref{eq:2D_DMoA_Solutions} contrasted with the BG eigenmodes of the finite lattice in (a-ii). (c) Localized eigenmodes of the corner, edge, and bulk defect resonances. ($m_\mathrm{r}=4, k_\mathrm{r}=1,k_\mathrm{g,r}=4,k_\mathrm{gd,r}=0$)}
    \label{fig:FIG5}
\end{figure*}

Now, consider the configurations described by the y-axis of the contour plots in Fig. \ref{fig:FIG4}a that represents a Multi-DMoA PnC with an edge ($x_\mathrm{d1}=0$) and a bulk ($x_\mathrm{d2}\in [1,\mathrm{N}-2]$) defect. Fig. \ref{fig:FIG4}b plots the numerical edge and bulk defect frequencies, $\Omega_\mathrm{d1}$ and $\Omega_\mathrm{d2}$, respectively, for different locations of the bulk defect, $x_\mathrm{d2}$ in the Multi-DMoA PnC. Consistent with the error plots, both $\Omega_\mathrm{d1}$ and $\Omega_\mathrm{d2}$ asymptote to the respective edge-bulk analytical estimates as the bulk defect migrates farther from the edge defect. To further explore interesting dynamics, we consider the Multi-DMoA PnC described by the \scalebox{1.25}{$\times$}-marker in Fig. \ref{fig:FIG4}a,b. Fig. \ref{fig:FIG4}c-i (Fig. \ref{fig:FIG4}c-ii) plots the FRF when the edge (bulk) defect is subject to a harmonic excitation and its displacement monitored. The relatively low amplitude and monotonic FRF curves when exciting the edge (bulk) defect mass at the other defect frequency, $\Omega_\mathrm{d2}$ ($\Omega_\mathrm{d1}$) presents an interesting insight in Fig. \ref{fig:FIG4}c-i (Fig. \ref{fig:FIG4}c-ii). The edge and bulk defect mode shapes plotted in Fig. \ref{fig:FIG4}c-iii explain this peculiar dynamical behavior of the Multi-DMoA PnC. In the bulk defect mode, $x_\mathrm{d2}=13$ constitutes an anti-node, while $x_\mathrm{d1}=0$ constitutes a node. Consequently, exciting the edge defect mass ($x_\mathrm{d1}=0$) at $\Omega_\mathrm{d2}$ fails to excite the bulk defect mode, and exciting the bulk defect mass ($x_\mathrm{d2}=13$) at $\Omega_\mathrm{d1}$ fails to excite the edge defect mode, leading to a relatively muted dynamical response at the respective excitation locations in both scenarios. Therefore, the analytical approach provides a robust estimate of the defect mode frequencies, mode shapes, and the velocity amplitude envelope even in the presence of multiple defects in the DMoA PnC, provided the additional defects are nodal points in the eigenmode associated with the considered defect.

\subsection{Defect modes in 2D DMoA PnCs}

This section proposes a solution approach to quantify the defect resonance frequency and mode shape in a 2D DMoA PnC. Consider the non-dimensional equations of motion of a uniform 2D membrane (Fig. \ref{fig:FIG5}(a-i)) modeled as a 2D MoA PnC:
\begin{equation}
        v_{,\bar{t}\bar{t}}^{(p,q)}+\left(4v^{(p,q)}-v^{(p+1,q)}-v^{(p-1,q)}-v^{(p,q+1)}-v^{(p,q-1)}\right)+k_\mathrm{g,r}v^{(p,q)}=0,
    \label{eq:2D_MoA_EOM}
\end{equation}
where, $v^{(p,q)}$ represents the out-of-plane displacement of the mass located in the $p-$th column and $q-$th row. The dispersion relations for the 2D membrane, assuming a plane wave \emph{ansatz}, $v^{(p,q)}=\tilde{v}^{(p,q)} e^{i(\mathbf{k}\cdot\mathbf{x}-\Omega \bar{t})} \ \forall \ \mathbf{k}=[\kappa_p,\kappa_q]^\mathrm{T}, \mathrm{x}=[x_p,x_q]^\mathrm{T}$, where $x_j=j-1$ for $j=p,q$, can be calculated as:
\begin{equation}            
\Omega(\mathbf{k})=\sqrt{k_\mathrm{g,r}+4\sin^2\left({\frac{\kappa_p}{2}}\right)+4\sin^2\left({\frac{\kappa_q}{2}}\right)}
    \label{eq:2D_Dispersion}
\end{equation}
where, $\mathrm{Re}\left[\kappa_{p,q}\right] \in[0,\pi]$. Fig. \ref{fig:FIG5}(b-i) plots the dispersion curve calculated in Eq. \eqref{eq:2D_Dispersion} using the primitive unit cell (Fig. \ref{fig:FIG5}(a-i)), associated with the irreducible Brillouin zone. 

Defect modes in 2D DMoA PnCs can be explored in three distinct configurations --  (i) a corner mode, (ii) an edge mode, and (iii) a bulk mode\cite{CaiJAM1995}. Though the direct application of the proposed analysis method (Eq. \eqref{eq:DMoA_MCE}) to 2D lattices can become cumbersome, we can develop a new method to analyze the defect modes in 2D DMoA PnCs taking insights from the 1D analysis.

 Consider the equations of motion of a defect mass in the 2D DMoA PnC:
\begin{equation}
            m_\mathrm{r}v_{,\bar{t}\bar{t}}^{(p,q)}+k_\mathrm{r}\left[n_cv^{(p,q)}-H\left(v^{(p',q')}\right)\right]+k_\mathrm{gd,r}v^{(p,q)}=0,
    \label{eq:2D_DMoA_EOM}
\end{equation}
where, $n_c$, and the function, $H$ represent the coordination number and the connectivity function, respectively, for different defect configurations listed in Tab. \ref{tab:2D_DMoA_EOM}. 

\begin{table}[H]
    \centering
    \begin{tabular}{|c|p{1.2cm}|p{1.2cm}|p{1.4cm}|p{1.2cm}|p{1.3cm}|p{1.3cm}|p{1.3cm}|p{1.2cm}|p{1.3cm}|}\hline
    \multirow{5}*{} & \multicolumn{9}{|c|}{Types of Defect}\\ \cline{2-10}
     & \multicolumn{4}{|c|}{Corner} & \multicolumn{4}{|c|}{Edge} & \multicolumn{1}{|c|}{Bulk} \\ \cline{2-10}
     & Left bottom, & Right bottom, & Right top, & Left top, & Bottom, & Right, & Top, & Left, & \\
     & $p=1$ & $p=\mathrm{N}$, & $p=\mathrm{N}$ & $p=1$, & $p\in\mathcal{S}$ & $p=\mathrm{N}$ & $p\in\mathcal{S}$, & $p=1$, & $p\in\mathcal{S}$\\
     & $q=1$ & $q=1$ & $q=\mathrm{N}$ & $q=\mathrm{N}$ & $q=1$ & $q\in\mathcal{S}$ & $q=\mathrm{N}$ & $q\in\mathcal{S}$ & $q\in\mathcal{S}$\\ \hline
     $n_c$ & 2 & 2 & 2 & 2 & 3 & 3 & 3 & 3 & 4\\ \hline
     \multirow{4}*{$H\left(v^{(p',q')}\right)$} & & & & & & & & & $v^{(p+1,q)}$\\ 
     & $v^{(1,2)}$ & $v^{(\mathrm{N}-1,1)}$ & $v^{(\mathrm{N}-1,\mathrm{N})}$ & $v^{(1,\mathrm{N}-1)}$ & $v^{(p+1,1)}$ & $v^{(\mathrm{N},q+1)}$ & $v^{(p+1,\mathrm{N})}$ & $v^{(1,q+1)}$ & $+v^{(p-1,q)}$\\
     & $+v^{(2,1)}$ & $+v^{(\mathrm{N},2)}$ & $+v^{(\mathrm{N},\mathrm{N}-1)}$ & $+v^{(2,\mathrm{N})}$ & $+v^{(p-1,1)}$ & $+v^{(\mathrm{N},q-1)}$ & $+v^{(p-1,\mathrm{N})}$ & $+v^{(1,q-1)}$ & $+v^{(p,q+1)}$\\
     & & & & & $+v^{(p,2)}$ & $+v^{(\mathrm{N}-1,q)}$ & $+v^{(p,\mathrm{N}-1)}$ & $+v^{(2,q)}$ & $+v^{(p,q-1)}$\\ \hline
    \end{tabular}
    \caption{Coordination number and interaction spring connections. $\mathcal{S}=[2,\mathrm{N}-1]$}
    \label{tab:2D_DMoA_EOM}
\end{table}

The continuum approximation for Eq. \eqref{eq:2D_MoA_EOM} yields:
\begin{equation*}
    v_{,\bar{t}\bar{t}}-\Delta v+k_\mathrm{g,r}v=0,
\end{equation*}
where, $\Delta=\frac{\partial^2}{\partial x_p^2}+\frac{\partial^2}{\partial x_q^2}$ is the Laplacian in the Cartesian plane, $x_p-x_q$. The 2D MoA PnC arrangement is such that, a single material defect at any lattice location will manifest an axis-symmetric decay. Assuming axis-symmetric conditions, the Laplacian can be transformed into polar coordinates: $\Delta=\frac{\partial^2}{\partial r^2}+\frac{1}{r}\cdot\frac{\partial}{\partial r}$. For a lattice location sufficiently far away, i.e., $r\rightarrow\infty$, from the defect location ($r=0$), the above equation becomes:
\begin{equation*}
    v_{,\bar{t}\bar{t}}-\frac{\partial^2 v}{\partial r^2}+k_\mathrm{g,r}v=0,
\end{equation*}
resembling the continuum approximation for a discrete 1D MoA PnC. Therefore, the axis-symmetric dispersion relation for waves traveling radially in a discrete lattice is given by Eq. \eqref{eq:MoA_Dispersion}.

From our 1D analysis in Sec. \ref{sec:DMoA}, we observe that the defect frequencies and decay rates, $\{\Omega_\mathrm{d},\kappa_I\}$ of 1D DMoA PnCs satisfy the dispersion relations of the underlying MoA PnC (Eq.\eqref{eq:MoA_Dispersion}, Fig. \ref{fig:FIG1}(a-i)). Informed by this observation, we combine the axis-symmetric dispersion (Eq. \eqref{eq:MoA_Dispersion}) and the equations of motion of the defect mass (Eq. \eqref{eq:2D_DMoA_EOM}), to predict the defect mode characteristics in 2D DMoA PnCs (Fig. \ref{fig:FIG5}(a-ii)). 

Given a radial decay rate, $\kappa_I$, the displacement of every mass connected to the defect mass can be given as: 
\begin{equation}
    v^{(p',q')}=v^{(p,q)}e^{-\kappa_I},
    \label{eq:2D_AxSymm}
\end{equation}
rendering $H\left(v^{(p',q')}\right)=n_cv^{(p,q)}e^{-\kappa_I}$. Substituting the \emph{ansatz} $v^{(p,q)}=\widetilde{v}_\mathrm{d}e^{i\Omega_\mathrm{d}\bar{t}}$ in Eq. \eqref{eq:2D_AxSymm} and combining it with Eq. \eqref{eq:MoA_Dispersion}, and Eq. \eqref{eq:2D_DMoA_EOM} , we get: 
\begin{equation}
    m_\mathrm{r}X+\frac{(m_\mathrm{r}-k_\mathrm{r}n_c)}{X}+2\delta=0,
    \label{eq:2D_DMoA_MCE}
\end{equation}
where,
\begin{equation*}
    X=e^{\kappa_I},\delta=\frac{(k_\mathrm{gd,r}-m_\mathrm{r}k_\mathrm{g,r})+(k_\mathrm{r}n_c-2m_\mathrm{r})}{2}
\end{equation*}
Rewriting Eq. \eqref{eq:2D_DMoA_MCE} in a quadratic form and estimating the roots, we get:
\begin{subequations}
    \begin{equation}
        \kappa_I=\log\left(\frac{-\delta+\sqrt{\delta^2-m_\mathrm{r}(m_\mathrm{r}-k_\mathrm{r}n_c)}}{m_\mathrm{r}}\right),
    \end{equation}
    \begin{equation}
        \Omega_\mathrm{d}=\sqrt{\frac{k_\mathrm{gd,r}+k_\mathrm{r}n_c(1-e^{-\kappa_I})}{m_\mathrm{r}}},
    \end{equation}
    \label{eq:2D_DMoA_Solutions}
\end{subequations}
Eq. \eqref{eq:2D_DMoA_Solutions} represents the analytical decay rates and the defect mode frequencies in a 2D DMoA PnC lattice with any of the three types of defects. 

Consider a 2D Multi-DMoA PnC ($\mathrm{N}_p=\mathrm{N}_q=21$) shown in Fig. \ref{fig:FIG5}(a-ii), embedded with a corner ($x_{p,q}=0$), an edge ($x_p=0,x_q=11$) and a bulk ($x_{p,q}=11$) defect. Applying Eq. \eqref{eq:2D_DMoA_Solutions}, we calculate the modal characteristics for each of the three defect resonances in the 2D lattice. Fig. \ref{fig:FIG5}(b-ii,iii) contrasts the analytical and numerical values of $\{\Omega_\mathrm{d},\kappa_I\}$ along both the in-plane ($x_{p,q}$) directions, and Fig. \ref{fig:FIG5}c plots the mode shapes of the respective BG resonances. The numerical and analytical $\Omega_\mathrm{d}$ values show an excellent agreement with an average error of $1.94\%$ between the three modes. The analytical decay rate also correlates well with the in-plane identical numerical decay rates, with a nominally higher average error of $5.23\%$ between the three modes. The analytical solutions have been tested for various material parameters and, show good correlation with their numerical counterparts with a low average error ($ \sim 1\%-8\%$). Hence, Eq. \eqref{eq:2D_DMoA_Solutions} presents a valuable analytical form of the modal characteristics of singular defects in a 2D DMoA PnC, that can be successfully extended to 2D Multi-DMoA PnCs (similar to the 1D Multi-DMoA PnC, Sec. \ref{sec:Multi-DMoA}) provided the embedded defects lie outside the proximal influence zones of each other.

\section{Applications of Defect-embedded PnCs}\label{sec:NumRes}

In this section, we demonstrate how tailored edge DMoA-based PnCs can -- (i) overcome the absence of an inertial frame of reference by accommodating an in-house virtual ground within the PnC structure, and (ii) manifest a customized amplitude and narrow-band frequency in response to broadband acoustic fields.

Engineering a phononic BG resonance\cite{HusseinPRSA2015,KianfarNJP2023} in PnCs has gained relevance in recent years in passive flow control applications, which brings together two rich fields of research - fluid dynamics, and PnCs and metamaterials. This presents an exciting opportunity for the two communities to collaborate and develop novel fluid-structure interaction strategies to achieve beneficial outcomes, e.g., fluid flow control using passive PnC designs, and efficiently harvesting energy from diverse fluid flows. In this context, the virtual ground design is useful in flow control applications where the PnC needs to be deployed in non-inertial environments, e.g., wings of moving aerial vehicles\cite{MachadoAIAASci2024}, underwater vehicles. An in-house virtual ground keeps the PnC isolated from external noise, e.g., wing vibrations, and engine noise, allowing effective airflow-PnC interaction leading to a significant reduction in operational and fuel costs by delaying laminar-to-turbulent transition\cite{HusseinPRSA2015}, reducing turbulent drag\cite{LinAPS2024}. Additionally, the relatively simple dynamical equations of DMoA-based PnCs and the existence of analytical relations between the defect mode characteristics and material properties, compared to other BG resonances (e.g., truncation resonances, topological modes) presents a favorable strategy to rapidly generate, test, and iterate PnC designs for flow control.

The customized acoustic absorber design is relevant for noise/vibration mitigation, and energy harvesting applications in environments subsumed in broadband acoustic noise, e.g., vehicular traffic, and machine noise in factories. The edge DMoA-based PnC can be customized to absorb a desirable amount of ambient energy at the desired frequency leading to improved operating environments.

Note we include nominal viscous damping, $c_\mathrm{m,r}=0.01$ and $c_\mathrm{k,r}=0$, in numerical simulations presented in this section, to realize bounded acoustic responses.

\subsection{PnC with a virtual ground}\label{sec:PnC_VG}

\begin{figure*}
    \includegraphics{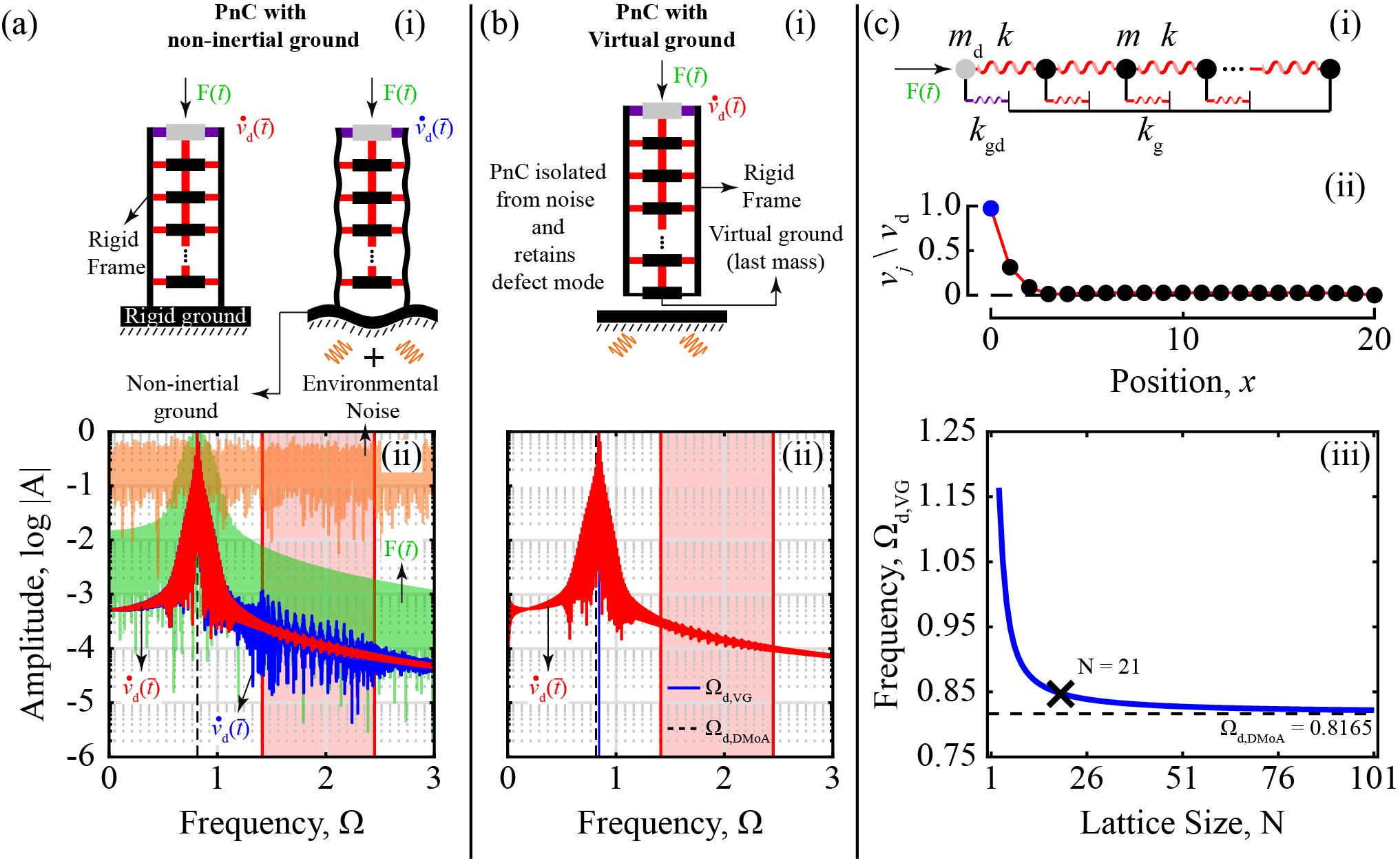}
    \caption{PnC with a virtual ground. (a-i) Schematics of PnCs featuring a rigid inertial and a non-inertial ground. (a-ii) FFTs of velocity response, $\dot{v}_\mathrm{d}(\bar{t})$ in an PnC with rigid inertial grounding (red) and non-inertial grounding (blue). Both PnCs are subject to force $\mathrm{F}(\bar{t})$ (green), however, the rigid grounding case has no environmental noise while the non-inertial grounding case features a broadband environmental noise (orange) contaminating the PnC response via the grounding springs. (b-i) Schematic of a modified PnC where the last mass functions as a virtual ground, and isolates the PnC from undesirable external noise while simultaneously retaining the edge defect mode characteristics relatively intact. (b-ii) FFTs of $\dot{v}_\mathrm{d}(\bar{t})$ in an PnC with virtual grounding subject to force $\mathrm{F}(\bar{t})$. (c-i) Reduced order mass-spring model of the PnC with virtual grounding, where grounding springs connect all masses to the last mass (opposite boundary), (c-ii) eigenmode of the PnC at $\Omega_\mathrm{d,VG}=0.8447$, and (c-iii) the convergence plot of the edge defect frequency, $\Omega_\mathrm{d,VG}$.}
    \label{fig:FIG6}
\end{figure*}

As described in Sec. \ref{sec:Model}, although grounding elements are not necessary for the existence of defect modes, their presence enables the existence of a defect mode as the principal eigenmode of a DMoA-based PnC within the bounded BG -- $\Omega\in[0,\sqrt{k_\mathrm{g,r}})$. However, PnCs and metamaterials have been deployed in a variety of environments, e.g., rotating shafts\cite{ArretcheMSSP2023}, aerial vehicles\cite{MachadoAIAASci2024}, seismic structures\cite{BrulePRL2014} where a true inertial frame of reference may not be accessible. In this context, we present a modified edge DMoA-based PnC that can function unhindered without a rigid inertial reference by utilizing one of its boundaries as a virtual ground. Though elastic cross-connections beyond nearest neighbors have been explored in periodic structures\cite{ChenNC2021,IglesiasSA2021} exhibiting rotons and statons, the edge defect-embedded PnC with a virtual ground presents a unique mechanically realizable design (see Fig. \ref{fig:SFIG4}) with simple inter-unit-cell connections, that has no infinitely-periodic counterpart to model the wave dispersion owing to the presence of material defects at a finite boundary.

The schematics in Fig. \ref{fig:FIG6}(a,b-i) contrast PnC configurations with an inertial, non-inertial, and virtual ground. Fig. \ref{fig:FIG6}(a,b-ii) contrasts the FFTs of the edge defect velocities when subject to a Gaussian wave-packet type forcing, $\mathrm{F}(\bar{t})$ (centered at $\Omega_\mathrm{d,DMoA}$) in these three grounding configurations. The FFT (red) in the rigid grounding scenario reveals a behavior consistent with dynamics characterized by the edge DMoA PnC model in Fig. \ref{fig:FIG1}b. However, the presence of broadband environmental noise (orange) renders the grounding mechanism non-inertial, causing the FFT (blue) in this scenario to deviate from its rigid grounding counterpart. Even though the environmental noise is chosen to have a maximum amplitude an order lower ($O(10^{-1})$) than the excitation signal, $\mathrm{F}(\bar{t})$, the impact on the PnC dynamics is evident.

Since the nature of the external grounding reference is dictated by the operating environment, we consider an alternate PnC configuration as shown in Fig. \ref{fig:FIG6}(b-i) to mitigate the impact of a non-inertial grounding reference. In the modified PnC, all constituent masses are now connected to the last mass which functions as an in-house virtual ground, isolating all PnC masses from the external environmental noise. Fig. \ref{fig:FIG6}(c-i) presents an equivalent mass-spring reduced order model for this altered PnC configuration with a virtual ground. An eigen-analysis (see Sec. \ref{sec:App_PnC_VG}) reveals completely new eigenvalues compared to the edge DMoA-based PnC, with three BG resonances -- $\Omega=0$ and $\Omega_\mathrm{d,VG}=0.8447$ in the low-frequency BG, and $\Omega=6.4095$ in the high-frequency BG, and the remainder lying within the PB range of the base MoA PnC. Resonances at $\Omega=0$ and $\Omega_\mathrm{d,VG}=0.8447$, respectively, correspond to the rigid body translation mode and the localized mode shape plotted in Fig. \ref{fig:FIG6}(c-ii) while the resonance at $\Omega=6.4095$ corresponds to a new eigenmode localized at the last mass, created as a result of the altered grounding spring connections. Though the eigenmode frequencies and shapes are significantly different from those of an edge DMoA-based PnC, the eigenmode at $\Omega_\mathrm{d,VG}=0.8447$ (Fig. \ref{fig:FIG6}(c-ii)) emerges as an exception, existing relatively unaltered compared to the edge defect mode shown in Fig. \ref{fig:FIG1}(b-ii), owing to the last mass being a nodal point in the defect mode. The FFT curve (red) in Fig. \ref{fig:FIG6}(b-ii) captures the PnC dynamics in the presence of an in-house virtual ground which is identical to that of an PnC with a rigid ground (Fig. \ref{fig:FIG6}(a-ii)). Therefore, our modified analytical approach in conjunction with the presence of an in-house virtual ground provides an enhanced capability to design a DMoA-based PnC that can function efficiently at a prescribed BG resonance regardless of the nature of the operating environment.

A closer look at Fig. \ref{fig:FIG6}(b-ii) reveals a slight change in defect modal frequency, $\Omega_\mathrm{d,VG}$ compared to the analytical DMoA PnC edge defect frequency, $\Omega_\mathrm{d,DMoA}=0.8165$. This discrepancy stems from the limitation that the decaying eigenmode reaches a near-zero and not an absolute zero displacement at the last mass, i.e., $v_{\mathrm{N}}\approx0$. Fig. \ref{fig:FIG6}(c-iii) illustrates this phenomenon, where choosing a larger PnC (large $\mathrm{N}$) aids in converging the displacement, $v_{\mathrm{N}}\rightarrow0$ and the frequency, $\Omega_\mathrm{d,VG}\rightarrow\Omega_\mathrm{d,DMoA}$.

A slight modification to the 3D structure described in Fig. \ref{fig:FIG1}(b-ii) can enable a physical realization of a PnC with a virtual ground. If the grounding frame is rigidly connected to the last aluminum plate (bottom boundary) of the 3D structure, instead of an external grounding reference (see Fig. \ref{fig:SFIG4}), the frame can essentially be viewed as an extension of the last aluminum plate allowing the grounding clips ($k_\mathrm{g}$), to elastically link the remaining plate masses to the virtual ground (last aluminum plate), as depicted in Fig. \ref{fig:FIG6}(b-ii).

\subsection{Customized acoustic absorber}
\begin{figure*}
    \centering
    \includegraphics{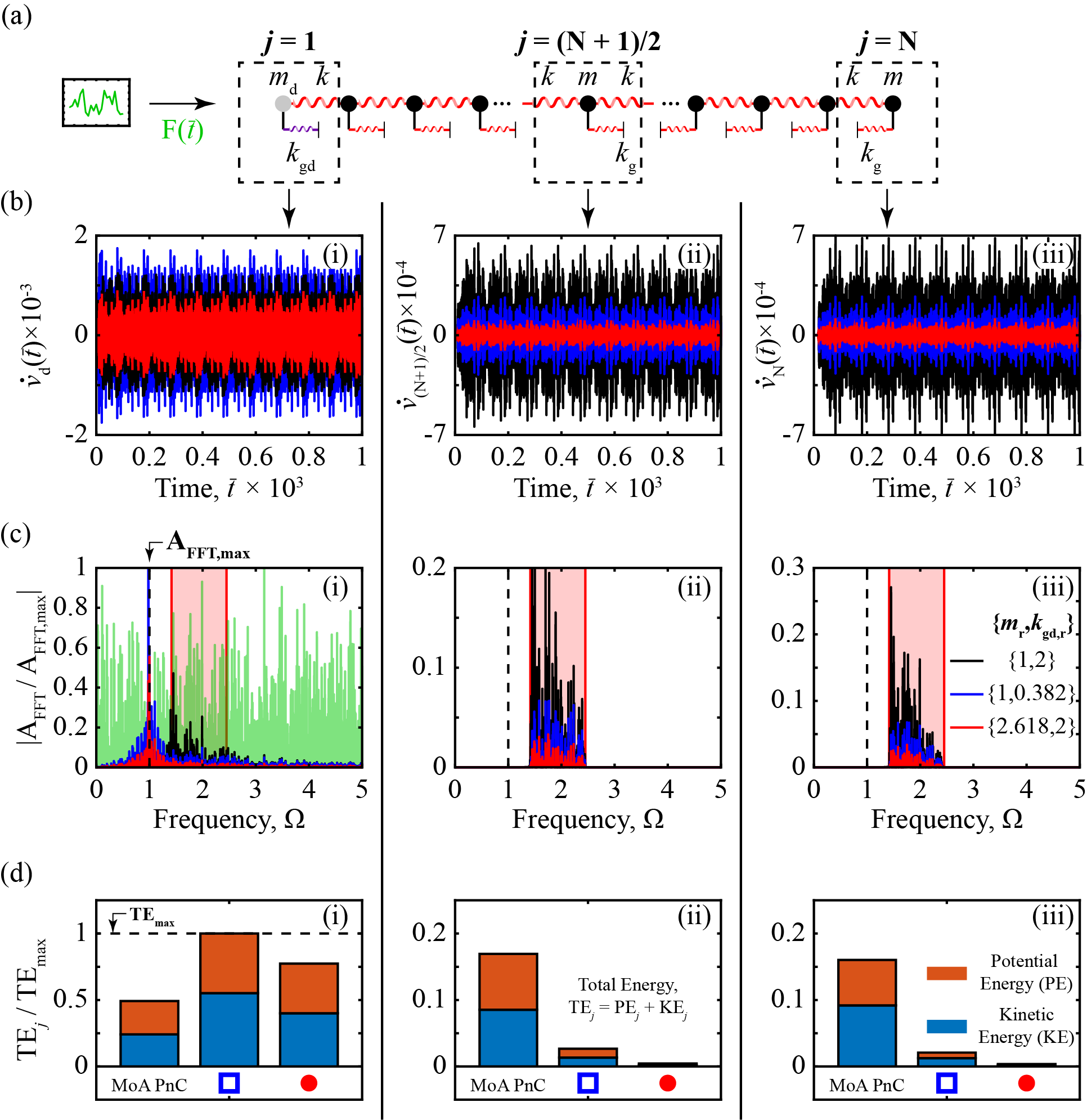}
    \caption{Customized acoustic absorber. (a) An edge DMoA-based PnC design with an engineered edge defect mode at $\Omega_\mathrm{d}=1$. (b) Velocity response, (c) the corresponding FFTs, and (d) Energy distribution at mass indices -- (i) $j=1$, (ii) $j=(\mathrm{N}+1)/2$, and (iii) $j=\mathrm{N}$ ($\mathrm{N}=51$), of three different PnC configurations - the base MoA PnC (black), and two edge DMoA PnCs with defect properties $\{m_\mathrm{r},k_\mathrm{gd,r}\}\rightarrow$ $\{1,0.382\}$ ($\square$), and $\{2.618,2\}$ ($\bullet$) used in Fig. \ref{fig:FIG3}.}
    \label{fig:FIG7}
\end{figure*}

Here we present an edge DMoA-based PnC that interacts with a broadband acoustic field, $\mathrm{F}(\bar{t})$, and manifests a prescribed narrow-band response at the edge defect while simultaneously customizing the transmission of PB frequencies through the PnC bulk. Fig. \ref{fig:FIG7}a shows the mass-spring configuration of an edge DMoA-based PnC subject to an external force, $\mathrm{F}(\bar{t})$ at the edge defect location. $\mathrm{F}(\bar{t})$ is chosen as a representative broadband acoustic signal (as illustrated by the green FFT curve in Fig. \ref{fig:FIG7}(c-i)) which can originate from a variety of environmental sources, e.g., vehicular traffic, machine noise in factories, fluid flow. A customized acoustic interaction of the PnC with this signal can be useful in mitigating or harvesting this ambient energy. To investigate this capability, we numerically simulate the response of the base MoA PnC (black) devoid of material defects and two DMoA PnC configurations described by edge defect parameters chosen from Fig. \ref{fig:FIG3} (blue and red). Fig. \ref{fig:FIG7}(b-d) plot, respectively, the velocity response of the three PnC configurations, the FFTs of these velocity signals (normalized by peak the largest FFT amplitude of all velocity signals), and the energy distribution at mass indices -- (i) $j=1$ (edge defect), (ii) $j=(\mathrm{N}+1)/2$ (midpoint), and (iii) $j=\mathrm{N}$ (last mass).

The velocity responses and the corresponding FFTs of the MoA PnC in Fig. \ref{fig:FIG7}(b,c) indicate its behavior as a passive acoustic filter that allows propagation of frequency components of $\mathrm{F}(\bar{t})$ in the PB range, $\Omega\in[\sqrt{k_\mathrm{g,r}},\sqrt{4+k_\mathrm{g,r}}]$ while attenuating frequencies in the BGs. Alternately, both the PnCs manifest a narrow-band velocity response, centered at the prescribed defect frequency, $\Omega_\mathrm{d}=1$ at $j=1$. Though PB frequency signatures of $\mathrm{F}(\bar{t})$ persist at $j=(\mathrm{N}+1)/2, \mathrm{N}$ in the PnCs, the level of transmission is significantly lower compared to the MoA PnC. To analyze this contrasting behavior, we quantify the total energy at a given mass index: 
\begin{equation*}
    \mathrm{TE}_j=\mathrm{KE}_j+\mathrm{PE}_j,
\end{equation*}
where the kinetic and potential energy components are calculated as:
\begin{equation*}
    \mathrm{KE}_j=\sum_{\bar{t}=0}^{\bar{t}=1000}\left\{
        \begin{array}{ll}
            \frac{1}{2}\left[m_\mathrm{d}\dot{v}_\mathrm{d}^2\right], & \mathrm{if} \ j=1,\\[0.3cm]
             \hfill \frac{1}{2}\left[m\dot{v}_j^2\right], & \mathrm{if} \ j \in [2,\mathrm{N}],
        \end{array}
    \right.     
\end{equation*}
\begin{equation*}
\mathrm{PE}_j=\sum_{\bar{t}=0}^{\bar{t}=1000}\left\{
        \begin{array}{ll}
            \frac{1}{2}\left[k_\mathrm{d}(v_\mathrm{d}-v_2)^2+k_\mathrm{gd}v_\mathrm{d}^2\right], & \mathrm{if} \ j=1,\\[0.3cm]
             \hfill \frac{1}{2}\left[k(v_j-v_{j+1})^2+k_\mathrm{g}v_j^2\right], & \mathrm{if} \ j \in [2,\mathrm{N-1}],\\[0.3cm]
             \hfill \frac{1}{2}\left[k_\mathrm{g}v_\mathrm{N}^2\right], & \mathrm{if} \ j=\mathrm{N},
        \end{array}
    \right.    
\end{equation*}
Consistent with the velocity responses in Fig. \ref{fig:FIG7}(b-ii,iii), the total energy bars, $\mathrm{TE}_{(\mathrm{N}+1)/2, \mathrm{N}}$ in Fig. \ref{fig:FIG7}(d-ii,iii), are significantly lower in both the PnCs compared to the MoA PnC indicating a reduction in transmission into the material bulk. This contrast in the transmission is facilitated by a superior energy localization at the edge defect in both PnCs, as evident from the higher energy content in Fig. \ref{fig:FIG7}(d-i). Furthermore, each PnC manifests a distinct energy localization behavior commensurate with the prescribed $\dot{\mathrm{V}}_\mathrm{E}$ scales from Fig. \ref{fig:FIG3}. The persistence of these prescribed velocity scales throughout the PnCs (Fig.\ref{fig:FIG7}(b,d-ii,iii)) reiterates the importance of the velocity amplitude envelope, $\dot{\mathrm{V}}_\mathrm{E}$ as a critical behavioral parameter that customizes both the localized response at the defect location and the energy transmission into the bulk. 

Additional modifications to the PnC, e.g., adding more defects\cite{JoIJMS2022}, secondary branching of the mass-spring PnC\cite{Deymier2013}, can be made according to operational requirements to further customize the PnC behavior without affecting the characteristics of the existing edge defect under reasonable proximity limits(\ref{sec:Multi-DMoA}). Therefore, our modified analytical method offers a systematic strategy to design multi-functional defect-embedded PnCs based on simple mass-spring reduced-order models.

\section{Conclusion}\label{sec:Conclusion}

In summary, we present a modified analytical approach that allows us to engineer BG resonances with prescribed characteristics in PnCs by embedding material defects and breaking the periodicity of the underlying homogeneous material. The perturbed Toeplitz matrix eigenvalue problem informs our analytical approach which accurately quantifies the acoustic characteristics of 1D and 2D DMoA PnCs with single and multiple defect-embedded with reasonable proximity limits. Furthermore, we demonstrate the benefits of defect modes through two DMoA-based PnC examples that feature an in-house virtual ground, and a customized acoustic absorption of an incident broadband acoustic field. The proposed strategy can be readily extended to lattices operating under general boundary conditions\cite{BastawrousJASA2022,HasanJEL2024} and polyatomic unit-cell\cite{HasanPRSA2019} based lattices, and study topologically protected modes\cite{HasanPRSA2019}, given at least a diatomic PnC. Therefore, the proposed method augments the toolkit for designing defect-based PnCs.

\section*{CRediT authorship contribution statement}
Vinod Ramakrishnan (VR): conceptualization, formal analysis, methodology, writing – original draft, review, and editing. Kathryn Matlack (KM): supervision, formal analysis, funding acquisition, writing – review, and editing. 

\section*{Data availability} 
Data will be made available on request. 

\section*{Acknowledgments} 
This material is based upon work supported by the Air Force Office of Scientific Research under award number FA9550-23-1-0299 and award number FA9550-21-1-0182. 

\section*{Appendix}

\setcounter{equation}{0}
\renewcommand{\theequation}{A\arabic{equation}}
\renewcommand{\theHequation}{A\arabic{equation}}
\setcounter{figure}{0}
\renewcommand{\thefigure}{A\arabic{figure}}
\renewcommand{\theHfigure}{A\arabic{figure}}
\setcounter{section}{0}
\renewcommand{\thesection}{A\arabic{section}}
\renewcommand{\theHsection}{A\arabic{section}}
\setcounter{table}{0}
\renewcommand{\thetable}{A\arabic{table}}
\renewcommand{\theHtable}{A\arabic{table}}

\section{Edge defect eigenmode}\label{sec:App_EdgeDefect}

The matrix equation of motion for an edge defect-embedded grounded monoatomic phononic crystal (edge DMoA PnC), with the temporal frequency normalized by $\omega_0=\sqrt{k/m}$, is given as:
\[\mathbf{M}\Ddot{\mathbf{V}}+\mathbf{K}\mathbf{V}=\mathbf{0}\]
where,
    \begin{equation}
            \mathbf{M}=\begin{bmatrix} 
            m_\mathrm{r} & 0 & \dots & 0\\
            0 & 1 & \dots & 0\\
            \vdots & \vdots & \ddots & \vdots\\
            0 & 0 & \dots & 1 
            \end{bmatrix}_{\mathrm{N}\times\mathrm{N}},
            \mathbf{K}=\begin{bmatrix}
            k_\mathrm{r}+k_\mathrm{gd,r} & -k_\mathrm{r} & 0 & 0 & \dots & 0\\
            -k_\mathrm{r} & 1+k_\mathrm{r}+k_\mathrm{g,r} & -1 & 0 & \dots & 0\\
            0 & -1 & 2+k_\mathrm{g,r} & -1 & \dots & 0\\
            0 & 0 & -1 & \ddots & \ddots & \vdots\\
            \vdots & \vdots & \vdots & \ddots &  & -1\\
            0 & 0 & 0 & \dots & -1 & 1+k_\mathrm{g,r}
            \end{bmatrix}_{\mathrm{N}\times\mathrm{N}},
            \label{eq:Matrices}
    \end{equation}
and, $m_\mathrm{r}=m_\mathrm{d}/m$, $k_\mathrm{r}=k_\mathrm{d}/k$, $k_\mathrm{g,r}=k_\mathrm{g}/k$ and $k_\mathrm{gd,r}=k_\mathrm{gd}/k$ represent the defect mass ratio, defect interaction stiffness ratio, grounding spring stiffness ratio and the defect grounding spring stiffness ratio, respectively. Note that defect eigenmodes exist for cases where, $k_\mathrm{r}\neq1$ however, the procedure to analytically estimate the eigenfrequencies is challenging when introducing a discrepancy in off-diagonal terms of the stiffness matrix. Hence, we set $k_r=1$ for all cases explored in this article to obtain a stiffness matrix conducive to formulating analytical solutions for the matrix eigenvalue problem.

Assuming harmonic motion, $\mathbf{V}=\hat{\mathbf{V}}e^{i\Omega \bar{t}}$, the eigenvalue problem for the DMoA PnC can be formulated as $|\mathbf{K}_\mathrm{d}|=\mathbf{0}$, where, 
\begin{equation}
\mathbf{K}_\mathrm{d}=\mathbf{K}-\Omega^2\mathbf{M}=
\begin{bmatrix} 
    d+\alpha & a & 0 & 0 & \dots & 0\\
    a & d & a & 0 & \dots & 0\\
    0 & a & d & a & \dots & 0\\
    0 & 0 & a & \ddots & \ddots & \vdots\\
    \vdots & \vdots & \vdots & \ddots & & a\\ 
    0 & 0 & 0 & \dots & a & d+\beta 
\end{bmatrix}
\label{eq:Matrix_EVP}
\end{equation}
is the dynamic stiffness matrix and the matrix coefficients are given as,
    \begin{equation*}
        d=2+k_\mathrm{g,r}-\Omega^2,a=-1, \beta=-1,
    \end{equation*}
    \begin{equation*}
        \alpha=(k_\mathrm{gd,r}-k_\mathrm{g,r})-1+(1-m_\mathrm{r})\Omega^2;
    \end{equation*}

The tridiagonal matrix, $\mathbf{K}_\mathrm{d}$ is consistent with the general form of a perturbed tridiagonal Toeplitz matrix\cite{FonsecaAMS2007,BastawrousJASA2022}. Without loss of generality, assuming an odd number of degrees of freedom in the system, i.e., $\mathrm{N}=2n+1$, the characteristic equation corresponding to an eigenvalue problem of a perturbed tridiagonal Toeplitz matrix is given by:
\begin{equation}
    (-d-\alpha-\beta)P_n-[\alpha\beta d + (\alpha+\beta)a^2]P_{n-1}=0
    \label{eq:DMoA_CE}
\end{equation}
where,
\[P_n=P_n(x)=a^{2n}U_n(x)\]
and, $U_n(x)$ is the Chebyshev polynomial of the second kind,
    \begin{equation*}
        U_n(x)=\frac{\sin(2(n+1)\kappa)}{\sin(2\kappa)},
    \end{equation*}
    \begin{equation*}
        x=\cos(2\kappa) = \frac{d^2}{2a^2}-1
    \end{equation*}
In the above equations, $\kappa$, represents the spatial frequency (non-dimensional wavenumber) associated with a given harmonic temporal frequency, $\Omega$. Eq. \eqref{eq:DMoA_CE} is an implicit equation that cannot be solved to obtain an explicit analytical form relating the DMoA PnC material parameters to $\kappa$ and $\Omega$. Hence, we rearrange the LHS in Eq. \eqref{eq:DMoA_CE} to generate the modified equation, Eq. \eqref{eq:DMoA_MCE} listed in the main text.

\begin{figure}
    \centering
    \includegraphics{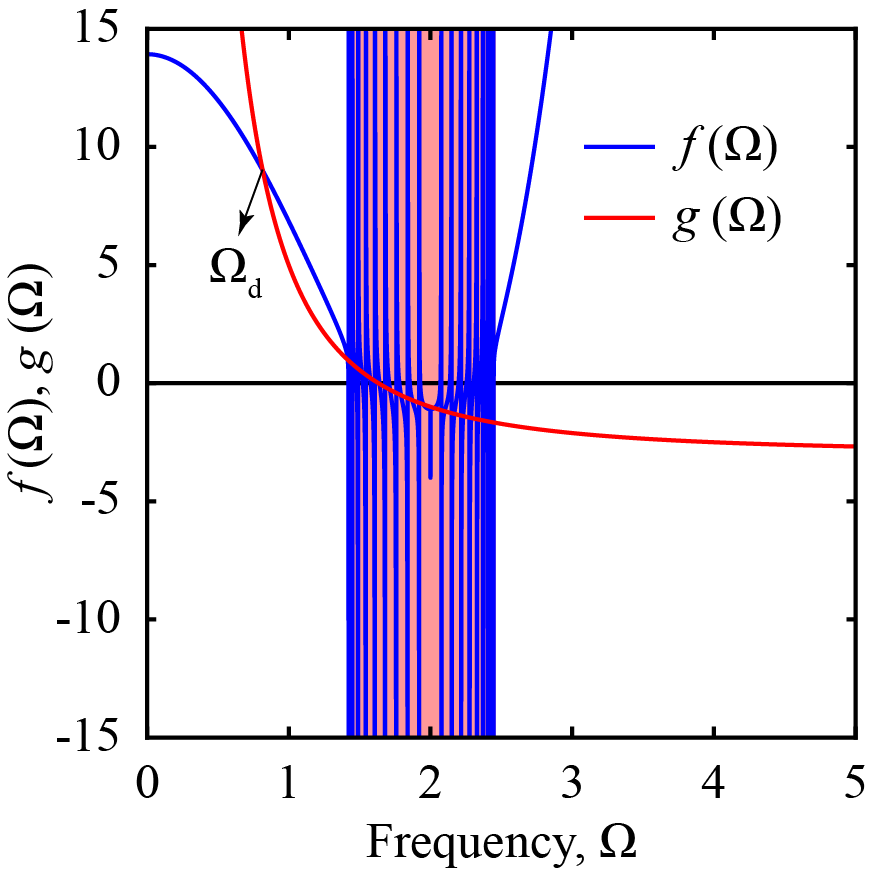}
    \caption{Graphical representation of functions $f(\Omega)$ and $g(\Omega)$ from Eq. \eqref{eq:DMoA_MCE} in the main text.}
    \label{fig:SFIG1}
\end{figure}

Eq. \eqref{eq:DMoA_MCE} is solved numerically by sweeping through a given range of frequencies, $\Omega$ to identify points of intersections between functions $f$ and $g$. A graphical representation is shown in Fig. \ref{fig:SFIG1}. Despite the U-transformation method utilized by Cai \emph{et al.}\cite{CaiJAM1995} for the case of a bulk defect and the perturbed Toeplitz matrix-based method that we propose for the edge defect being fundamentally different solution approaches, a noteworthy similarity between the procedures is that they both involve implicit equations in the final step that need to be solved numerically to obtain the defect eigenfrequencies and eigenmode shapes.

\subsection*{Damped DMoA PnCs}
The normalized matrix equation of motion for a damped DMoA PnC is given as:
\[\mathbf{M}\ddot{\mathbf{V}}+\mathbf{C}\dot{\mathbf{V}}+\mathbf{K}\mathbf{V}=\mathbf{0}\]
where, the general damping matrix has the form, 
    \begin{equation}
            \mathbf{C}=c_\mathrm{on,r}\mathbf{I}+c_\mathrm{in,r}\mathbf{K}_\mathrm{in}=\begin{bmatrix}
            c_\mathrm{on,r}+c_\mathrm{in,r} & -c_\mathrm{in,r} & 0  & \dots & 0\\
            -c_\mathrm{in,r} & c_\mathrm{on,r}+2c_\mathrm{in,r} & -c_\mathrm{in,r} & \dots & 0\\
            0 & -c_\mathrm{in,r}  & \ddots & \ddots & \vdots\\
            \vdots & \vdots & \vdots & \ddots  & -c_\mathrm{in,r}\\
            0 & 0 & \dots & -c_\mathrm{in,r} & c_\mathrm{on,r}+c_\mathrm{in,r}
            \end{bmatrix}_{\mathrm{N}\times\mathrm{N}},
            \label{eq:Damping_Matrix}
    \end{equation}
\begin{equation*}
    \mathbf{I}=\begin{bmatrix}
        1 & 0 & \cdots & 0\\
        0 & 1 & \cdots & 0\\
        \vdots & \vdots & \ddots & \vdots\\
        0 & 0 & \cdots & 1
    \end{bmatrix}_{\mathrm{N}\times\mathrm{N}},
    \mathbf{K}_\mathrm{in}=\begin{bmatrix}
        1 & -1 & 0 & \cdots & 0\\
        -1 & 2 & -1 &\cdots & 0\\
        0 & -1 & & \ddots & \vdots\\
        \vdots & \vdots & \ddots & \ddots & -1\\
        0 & 0 & \cdots & -1 & 1
    \end{bmatrix}_{\mathrm{N}\times\mathrm{N}},
\end{equation*}
Consequently, the dynamic stiffness matrix and matrix coefficients can be modified as - $\mathbf{K}_\mathrm{d}=\mathbf{K}+i\Omega\mathbf{C}-\Omega^2\mathbf{M}$ 
    \begin{equation*}
        d=2+k_\mathrm{g,r}+i\Omega(c_\mathrm{on,r}+2c_\mathrm{in,r})-\Omega^2,a=-(1+i\Omega c_\mathrm{in,r}), \beta=-(1+i\Omega c_\mathrm{in,r}),
    \end{equation*}
    \begin{equation*}
        \alpha=(k_\mathrm{gd,r}-k_\mathrm{g,r})-(1+i\Omega c_\mathrm{in,r})+(1-m_\mathrm{r})\Omega^2
    \end{equation*}

\section{Multiple BG resonances with single edge defect}\label{sec:App_MultiBGRes}
\begin{figure*}
    \centering
    \includegraphics{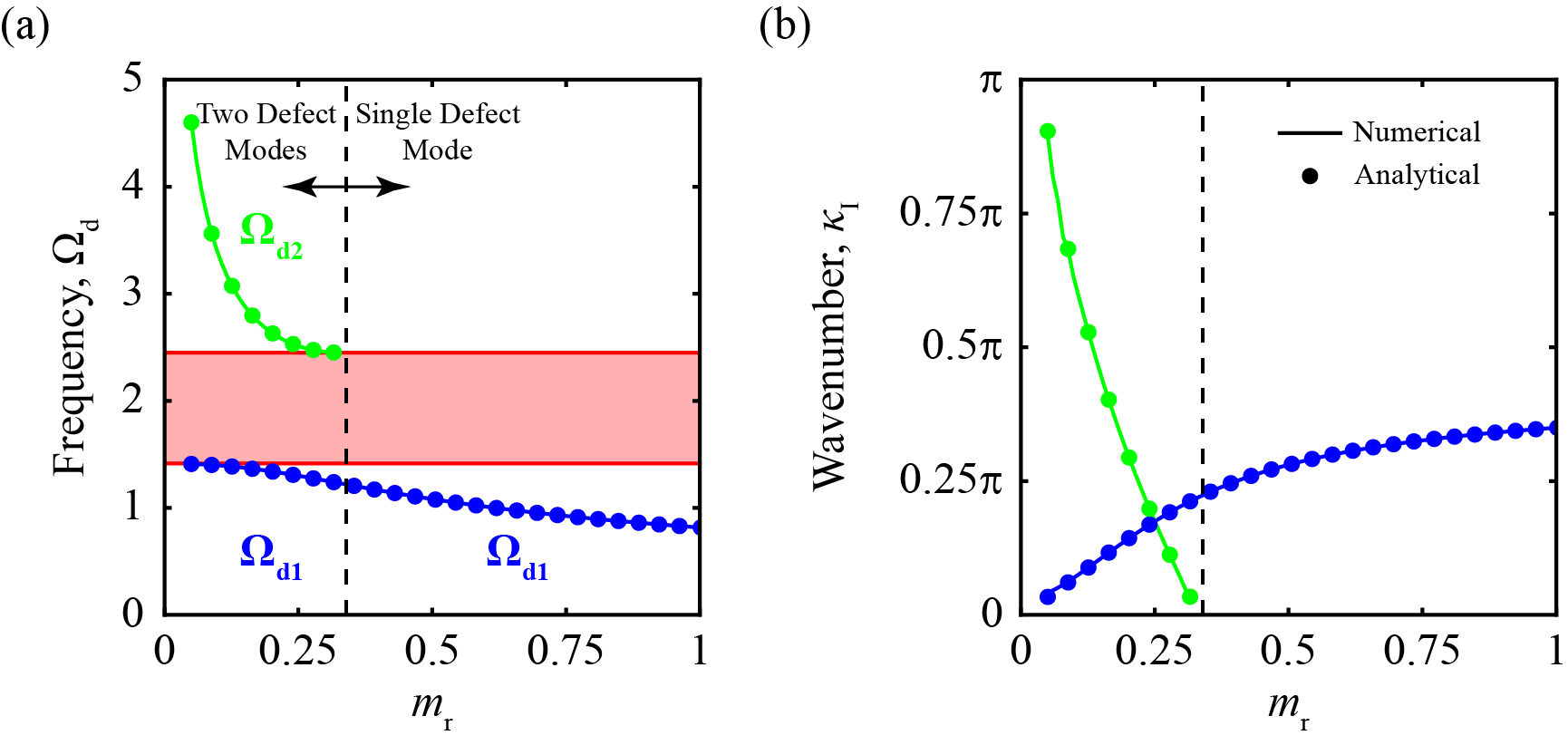}
    \caption{Transition of a DMoA PnC with material defect at a single boundary from manifesting a single defect mode to two defect modes. (a) Defect frequencies, $\Omega_\mathrm{d1,d2}$ and, the corresponding (b) decay rates, $\kappa_I$. ($k_\mathrm{gd,r}=0$)}
    \label{fig:SFIG2}
\end{figure*}

Though, the design space of the defect properties, $\{m_\mathrm{r},k_\mathrm{gd,r}\}$ probed in Fig. \ref{fig:FIG2} in the main text manifest a single defect mode, other combinations, especially cases where $m_\mathrm{r}<1$ and $k_\mathrm{gd,r}<k_\mathrm{g,r}$ can manifest an additional defect mode in $2^\mathrm{nd}$ BG of the MoA PnC, localized at the same edge defect location as the existing defect mode. Fig. \ref{fig:SFIG2} presents the (a) defect frequencies, $\Omega_\mathrm{d1,d2}$ and (b) the decay rates, $\kappa_I$ when probing the defect material properties, $m_\mathrm{r}\in[0.04,1]$ and $k_\mathrm{gd,r}=0$. Not only do the analytical and numerical values of the defect frequencies and decay rates agree well, but we also notice that below the value, $m_\mathrm{r}\approx0.33$ the eigenfrequencies of the finite DMoA PnC undergo a bifurcation, where the highest PB frequency relocates into the $2\mathrm{nd}$ BG, appearing as a second defect resonance frequency in the DMoA PnC. 

\section{Resonance Velocity Amplitude Envelope}\label{sec:RVAE}

The matrix equation governing the forced response of a DMoA PnC subject to an external, single-frequency harmonic forcing is given as:
\begin{equation}
    \mathbf{M}\Ddot{\mathbf{V}}+\mathbf{K}\mathbf{V}=\frac{\mathbf{F}_0}{2}e^{i\Omega_\mathrm{exc} \bar{t}} +\mathrm{c.c.},
    \label{eq:Forced_EOM}
\end{equation}
where, $\mathbf{F}_0$ represents an $\mathrm{N}\times 1$ column vector with complex force excitation amplitudes (i.e., $\mathbf{F}_{0,j} \in \mathbb{C}$ for $j=1,2,\cdots,\mathrm{N}$), $\Omega_\mathrm{exc}$ represents the non-dimensional excitation frequency, and $\mathrm{c.c.}$ stands for the complex conjugate terms. If the excitation frequency coincides with one of the natural frequencies of the systems, i.e., $\Omega_\mathrm{exc}=\Omega_j$, the general solution \emph{ansatz} for the displacement of the DMoA PnC at resonance is of the form:
\begin{equation}
    \mathbf{V}(\bar{t})=\hat{\mathbf{V}}_j\left(\frac{A+B\bar{t}}{2}\right)e^{i\Omega_j \bar{t}} + \mathrm{c.c.},
    \label{eq:Forced_EOM_Soln1}
\end{equation}
where, $\hat{\mathbf{V}}_j$ represents the normalized eigenvector (i.e., $\hat{\mathbf{V}}_j^\mathrm{T}\hat{\mathbf{V}}_j=1$) corresponding to $\Omega_j$ and coefficients $ A, B\in \mathbb{C}$. Specifically, for $\Omega_j=\Omega_\mathrm{d}$ and, initial displacement and velocity conditions - $\mathbf{V}(0)=\mathbf{0}$ and $\dot{\mathbf{V}}(0)=\mathbf{0}$, the eigenvector, $\hat{\mathbf{V}}_\mathrm{d}$ and the coefficients $A,B$ can be written as:
\begin{subequations}
    \begin{equation}
        \hat{\mathbf{V}}_\mathrm{d}=\widetilde{\mathrm{V}}_\mathrm{d}\mathbf{V}_\mathrm{d}=\widetilde{\mathrm{V}}_\mathrm{d}e^{i\kappa_R \mathbf{x}}e^{-\kappa_I\mathbf{x}},
    \end{equation}
\begin{align}
    \mathbf{V}(0)=\mathbf{0} \Longrightarrow & \mathrm{Re}(A)=0 \nonumber\\
    \Longrightarrow & A=ia, \mathrm{for} \ a \in \mathbb{R},
\end{align}
\begin{align}
    \dot{\mathbf{V}}(0)=\mathbf{0} \Longrightarrow & \mathrm{Re}(B)=\Omega_\mathrm{d}\mathrm{Im}(A) \nonumber\\
    \Longrightarrow & B=a\Omega_\mathrm{d}+ib, \mathrm{for} \ a,b \in \mathbb{R}, 
\end{align}
\label{eq:Forced_EOM_Soln2}
\end{subequations}
where, $\mathbf{x}=[0,1,\cdots,\mathrm{N}-1]^\mathrm{T}$ represents the coordinate vector, $\kappa_R$ and $\kappa_I$, respectively, represent the real and imaginary wavenumber (decay rate), and $\widetilde{\mathrm{V}}_\mathrm{d}$ represents the normalizing factor associated with the defect eigenmode, $\hat{\mathbf{V}}_\mathrm{d}$. Assuming a defect mode in the $1^\mathrm{st}$ BG ($2^\mathrm{nd}$ BG), i.e., $\kappa_R=0$ ($\kappa_R=\pi$) and, substituting, Eq. \eqref{eq:Forced_EOM_Soln1} and Eq. \eqref{eq:Forced_EOM_Soln2} in Eq. \eqref{eq:Forced_EOM}, we get:

\begin{equation}
    \left\{\left(\frac{\Omega_\mathrm{d}}{2}\right)\left[a\bar{t}(\mathbf{K}-\Omega_\mathrm{d}^2\mathbf{M})-2b\mathbf{M}\right]+
    \left(\frac{i}{2}\right)\left[\mathbf{K}(a+b\bar{t})+\mathbf{M}(a-b\bar{t})\Omega_\mathrm{d}^2\right]\right\}\hat{\mathbf{V}}_\mathrm{d}e^{i\Omega_\mathrm{d} \bar{t}} + \mathrm{c.c.} = \frac{\mathbf{F}_0}{2}e^{i\Omega_\mathrm{d} \bar{t}} +\mathrm{c.c.},
    \label{eq:Resonance_EOM1}
\end{equation}

The real and imaginary parts of the coefficients of $e^{i\Omega_\mathrm{d} \bar{t}}$ on the LHS and RHS in Eq. \eqref{eq:Resonance_EOM1} can be equated to obtain two matrix equations:
\begin{subequations}
    \begin{equation}
        \left[a\bar{t}(\mathbf{K}-\Omega_\mathrm{d}^2\mathbf{M})-2b\mathbf{M}\right]\hat{\mathbf{V}}_\mathrm{d}=\frac{\mathrm{Re}(\mathbf{F}_0)}{\Omega_\mathrm{d}};
    \end{equation}
    \begin{equation}
        \left[\mathbf{K}(a+b\bar{t})+\mathbf{M}(a-b\bar{t})\Omega_\mathrm{d}^2\right]\hat{\mathbf{V}}_\mathrm{d}=\mathrm{Im}(\mathbf{F}_0),
    \end{equation}
    \label{eq:Resonance_EOM2}
\end{subequations}

Now, assume an external sinusoidal forcing, $\mathbf{F}=[\mathrm{F}_0\sin (\Omega_\mathrm{d} \bar{t}),0,\cdots,0]^\mathrm{T}$, applied on the DMoA PnC boundary housing the (edge) defect. The complex forcing amplitudes are expressed as:
\begin{equation*}
    \mathrm{Re}(\mathbf{F}_0)=\mathbf{0}, \mathrm{Im}(\mathbf{F}_0)=[-\mathrm{F}_0,0,\cdots,0]^\mathrm{T},
\end{equation*}
Substituting, the above equation into Eq. \eqref{eq:Resonance_EOM2}a, we get $b=0$, as $\Omega_\mathrm{d}$ and $\hat{\mathbf{V}}_\mathrm{d}$ are a solution to the eigenvalue problem $\mathbf{K}_\mathrm{d}\hat{\mathbf{V}}_\mathrm{d}=0$. Therefore, the DMoA PnC response to the given sinusoidal forcing can be written as:
\begin{align}
    \mathbf{V} &= \left(\frac{\dot{\mathrm{V}}_\mathrm{E}}{\Omega_\mathrm{d}^2}\right)\left[\sin(\Omega_\mathrm{d}\bar{t})-\Omega_\mathrm{d}\bar{t}\cos(\Omega_\mathrm{d}\bar{t})\right]\mathbf{V}_\mathrm{d},\\ \nonumber 
    \dot{\mathbf{V}} &= \left[\dot{\mathrm{V}}_\mathrm{E}\bar{t}\sin(\Omega_\mathrm{d}\bar{t})\right]\mathbf{V}_\mathrm{d},
\end{align}
where, Eq. \eqref{eq:Resonance_EOM2}b yields:
\begin{equation}
\dot{\mathrm{V}}_\mathrm{E}=-\Omega_\mathrm{d}^2\widetilde{\mathrm{V}}_\mathrm{d}\left[\hat{\mathbf{V}}_\mathrm{d}^\mathrm{T}\left[\mathbf{K}+\mathbf{M}\Omega_\mathrm{d}^2\right]^{-1}\mathrm{Im}(\mathbf{F}_0)\right],
    \label{eq:Resonance_Coeff1}
\end{equation}

\section{Steady-State Velocity Amplitude}\label{sec:App_SS_VelAmp}
The matrix equation governing the forced response of a damped DMoA PnC can be written as:
\begin{equation}
    \mathbf{M}\ddot{\mathbf{V}}+\mathbf{C}\dot{\mathbf{V}}+\mathbf{K}\mathbf{V}=\frac{\mathbf{F}_0}{2}e^{i\Omega_\mathrm{exc} \bar{t}} +\mathrm{c.c.},
    \label{eq:Damp_Forced_EOM}
\end{equation}
The general \emph{ansatz} for the displacement of the DMoA PnC is of the form:
\begin{equation}
    \mathbf{V}(\bar{t})=\sum _{j=1}^N \left(\frac{A_j}{2}\right) e^{(-\xi_j\Omega_{n,j}+i\Omega_j )\bar{t}}\hat{\mathbf{V}}_j +\left(\frac{B_\mathrm{exc}}{2}\right) e^{i\Omega_j \bar{t}}\hat{\mathbf{V}}_\mathrm{exc}+ \mathrm{c.c.},
    \label{eq:DForced_EOM_Soln1}
\end{equation}
where, $\xi_j$  is the modal damping ratio, $\Omega_j$ is the $j^\mathrm{th}$ damped natural frequency, $\Omega_{n,j}=\left|\Omega_j/\sqrt{1-\xi_j^2}\right|$, and coefficients $ A_j, B\in \mathbb{C}$. When exciting at the defect resonance frequency, $\Omega_j=\Omega_\mathrm{d}$ the coefficients $A_{j}=0 \ \forall \ j\neq \mathrm{d}$. Applying initial displacement and velocity conditions - $\mathbf{V}(0)=\mathbf{0}$ and $\dot{\mathbf{V}}(0)=\mathbf{0}$, assuming, $A_\mathrm{d}=a_R+ia_I$ and $B_\mathrm{exc}=B_\mathrm{d}=b_R+ib_I$:
\begin{subequations}
\begin{align}
    \mathbf{V}(0)=\mathbf{0} \Longrightarrow & \mathrm{Re}(A_\mathrm{d}+B_\mathrm{d})=0 \nonumber\\
    \Longrightarrow & a_R=-b_R, \mathrm{for} \ a_R \in \mathbb{R},
\end{align}
\begin{align}
    \dot{\mathbf{V}}(0)=\mathbf{0} \Longrightarrow & \xi_\mathrm{d}\Omega_{n,\mathrm{d}}\mathrm{Re}(A_\mathrm{d})=\Omega_\mathrm{d}\mathrm{Im}(A_\mathrm{d}+B_\mathrm{d}) \nonumber\\
    \Longrightarrow & \xi_\mathrm{d}\Omega_{n,\mathrm{d}}a_R=\Omega_\mathrm{d}(a_I+b_I) , \mathrm{for} \ a_I,b_I \in \mathbb{R}, 
\end{align}
\label{eq:DForced_EOM_Soln2}
\end{subequations}
Substituting the \emph{ansatz} -  $\mathbf{V}(\bar{t})=\frac{A_\mathrm{d}}{2}e^{(-\xi_\mathrm{d}\Omega_{n,\mathrm{d}}+i\Omega_\mathrm{d} )\bar{t}}\hat{\mathbf{V}}_\mathrm{d} +\frac{B_\mathrm{d}}{2} e^{i\Omega_\mathrm{d} \bar{t}}\hat{\mathbf{V}}_\mathrm{d}+ \mathrm{c.c.},$ into Eq. \eqref{eq:Damp_Forced_EOM}:
\begin{equation}
\frac{B_\mathrm{d}}{2}\left(-\Omega_\mathrm{d}^2\mathbf{M}+i\Omega_\mathrm{d}\mathbf{C}+\mathbf{K}\right)\hat{\mathbf{V}}_\mathrm{d}e^{i\Omega_\mathrm{d}\bar{t}}+\mathrm{c.c.}=\frac{\mathbf{F}_0}{2}e^{i\Omega_\mathrm{d} \bar{t}} +\mathrm{c.c.},
\label{eq:Damped_Resonance_EOM1}
\end{equation}
The real and imaginary parts of the coefficients of $e^{i\Omega_\mathrm{d} \bar{t}}$ on the LHS and RHS in Eq. \eqref{eq:Damped_Resonance_EOM1} can be equated to obtain two matrix equations:
\begin{subequations}
    \begin{equation}
\left[-b_R\Omega_\mathrm{d}^2\mathbf{M}-b_I\Omega_\mathrm{d}\mathbf{C}+b_R\mathbf{K}\right]\hat{\mathbf{V}}_\mathrm{d}=\mathrm{Re}(\mathbf{F}_0);
    \end{equation}
    \begin{equation}
        \left[-b_I\Omega_\mathrm{d}^2\mathbf{M}+b_R\Omega_\mathrm{d}\mathbf{C}+b_I\mathbf{K}\right]\hat{\mathbf{V}}_\mathrm{d}=\mathrm{Im}(\mathbf{F}_0),
    \end{equation}
    \label{eq:Damped_Resonance_EOM2}
\end{subequations}
For a sinusoidal forcing ($\mathrm{Re}(\mathbf{F}_0)=0$), we get:
    \begin{equation*}
a_R=-b_R=-b_I\Omega_\mathrm{d}\alpha,
    \end{equation*}
    where, $\alpha=\hat{\mathbf{V}}_\mathrm{d}^\mathrm{T}\left[\mathbf{K}-\Omega_\mathrm{d}^2\mathbf{M}\right]^{-1}\mathbf{C}\hat{\mathbf{V}}_\mathrm{d}$, and,
    \begin{equation*}
        a_I=\frac{\xi_\mathrm{d}\Omega_{n,\mathrm{d}}a_R}{\Omega_\mathrm{d}}-b_I\Longrightarrow a_I=-b_I(1+\xi_\mathrm{d}\Omega_{n,\mathrm{d}}\alpha),
    \end{equation*}
    \begin{equation*}
        b_I=\left[\frac{\alpha}{1+\alpha^2\Omega_\mathrm{d}^2}\right]\hat{\mathbf{V}}_\mathrm{d}^\mathrm{T}\mathbf{C}^{-1}\mathrm{Im}(\mathbf{F}_0),
    \end{equation*}
Therefore, the analytical form of the DMoA PnC response is given as:
\begin{equation*}
    \mathbf{V}(\bar{t})=-\left(\frac{b_I}{2}\right)\left[\alpha\Omega_\mathrm{d}+i(1+\xi_\mathrm{d}\Omega_\mathrm{n,d}\alpha)\right]e^{(-\xi_\mathrm{d}\Omega_\mathrm{n,d}+i\Omega_\mathrm{d}) \bar{t}}\hat{\mathbf{V}}_\mathrm{d}+\left(\frac{b_I}{2}\right)(i+\alpha\Omega_\mathrm{d}) e^{i\Omega_\mathrm{d} \bar{t}}\hat{\mathbf{V}}_\mathrm{d}+ \mathrm{c.c.},
\end{equation*}
that, yields the steady state ($\bar{t}\rightarrow\infty$) response:
\begin{align}
\mathbf{V}(\bar{t})=&\left(\frac{b_I}{2}\right)(i+\alpha\Omega_\mathrm{d}) e^{i\Omega_\mathrm{d} \bar{t}}\hat{\mathbf{V}}_\mathrm{d}+ \mathrm{c.c.}\nonumber\\
=& b_I\left[\alpha\Omega_\mathrm{d}\cos(\Omega_\mathrm{d}\bar{t})-\sin(\Omega_\mathrm{d}\bar{t})\right]\hat{\mathbf{V}}_\mathrm{d}\nonumber\\
\dot{\mathbf{V}}=&-b_I\Omega_\mathrm{d}\left[\alpha\Omega_\mathrm{d}\sin(\Omega_\mathrm{d}\bar{t})+\cos(\Omega_\mathrm{d}\bar{t})\right]\hat{\mathbf{V}}_\mathrm{d}
\end{align}

\section{Multiple defects}\label{sec:App_MultiDef}
\begin{figure*}
    \centering
    \includegraphics{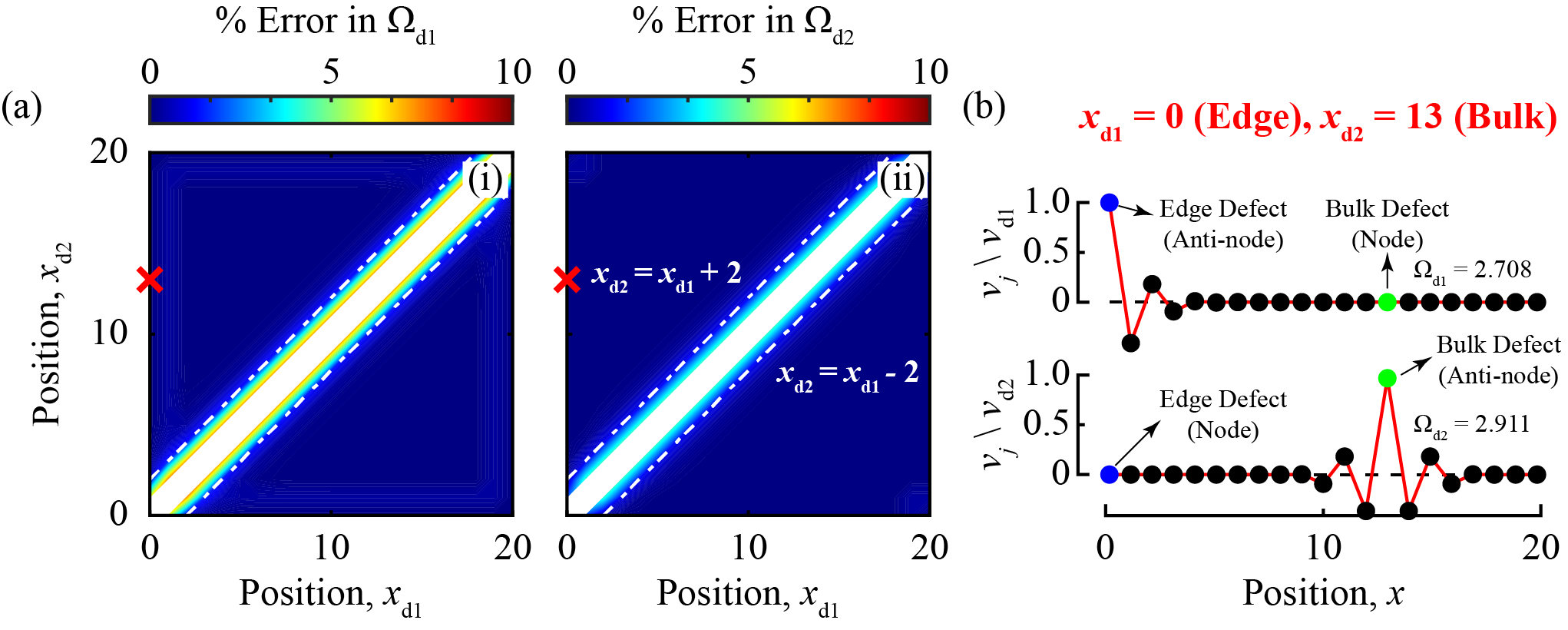}
    \caption{1D Multi-DMoA PnC with defect frequencies in $2^\mathrm{nd}$ BG of the MoA PnC. (a) Absolute percentage error contours for a Multi-DMoA PnC ($\mathrm{N}=21$) with two defects at $x_\mathrm{d1}, x_\mathrm{d2} \in [0,\mathrm{N}-1]$ and $x_\mathrm{d1} \neq x_\mathrm{d2}$, (b) Eigenmode shapes of the Multi-DMoA PnC with one edge ($x_\mathrm{d1}=0$) and one bulk ($x_\mathrm{d2}$) defect. Multi-DMoA PnC material parameters - $m=1$, $k=1$, $k_\mathrm{g,r}=2$, $k_\mathrm{r}=1$ and $\{m_r,k_\mathrm{gd,r}\}=\{1,6\}$}
    \label{fig:SFIG3}
\end{figure*}

The modified analytical approach presented in this article can be applied to determine the defect resonance frequencies of a Multi-DMoA PnC provided the additional defects remain relatively dormant at the resonance frequency of a considered defect. To verify the accuracy of the analytical estimate of $\Omega_\mathrm{d}$, Fig. \ref{fig:FIG4} in the main text plots the absolute percentage error b.w. the numerical and analytical values of the defect frequency given by,
\begin{equation}
    \% \ \mathrm{Error} \  \mathrm{in} \ \Omega_{\mathrm{d}}=\frac{\Delta \Omega_{\mathrm{d}}}{\Omega_{\mathrm{d},\mathrm{An.}}}\times 100\%,
    \label{eq:AP_Error}
\end{equation}
where, $\Delta \Omega_{\mathrm{d}}=\left|\Omega_{\mathrm{d},\mathrm{Num.}}-\Omega_{\mathrm{d},\mathrm{An.}}\right|$. 

When the defect frequencies lie in the $1^\mathrm{st}$ BG of the MoA PnC, we notice a small surge in the $\Omega_\mathrm{d1}$ error when at least one of the penultimate masses are prescribed defects, i.e., $x_\mathrm{d1,d2}=1 \ \mathrm{or} \ \mathrm{N}-2$. This error surge is attributed to the spatial in-phase vibration associated with the defect modes in the $1^\mathrm{st}$ BG and the significant impact of defect proximity to the finite boundary. Alternately, Fig. \ref{fig:SFIG3} explores a Multi-DMoA PnC with defect modes now in the $2^\mathrm{nd}$ BG of a MoA PnC. The error contours at $x_\mathrm{d1,d2}=1 \ \mathrm{or} \ \mathrm{N}-2$ in Fig. \ref{fig:SFIG3}a are virtually indistinguishable from the surrounding configurations (blue region - $0.003\%$, surge region - $0.29\%$) as opposed to the error contours in Fig. \ref{fig:FIG4}a (blue region - $0.004\%$, surge region - $3.9\%$). The reduction in the error surge is indeed a result of improved accuracy of the analytical model and out-of-phase vibration of the defect and boundary masses as the absolute error, $\Delta\Omega_{\mathrm{d}}$ also decreases for defect modes in the $2^\mathrm{nd}$ BG.

\begin{figure*}
    \centering
    \includegraphics{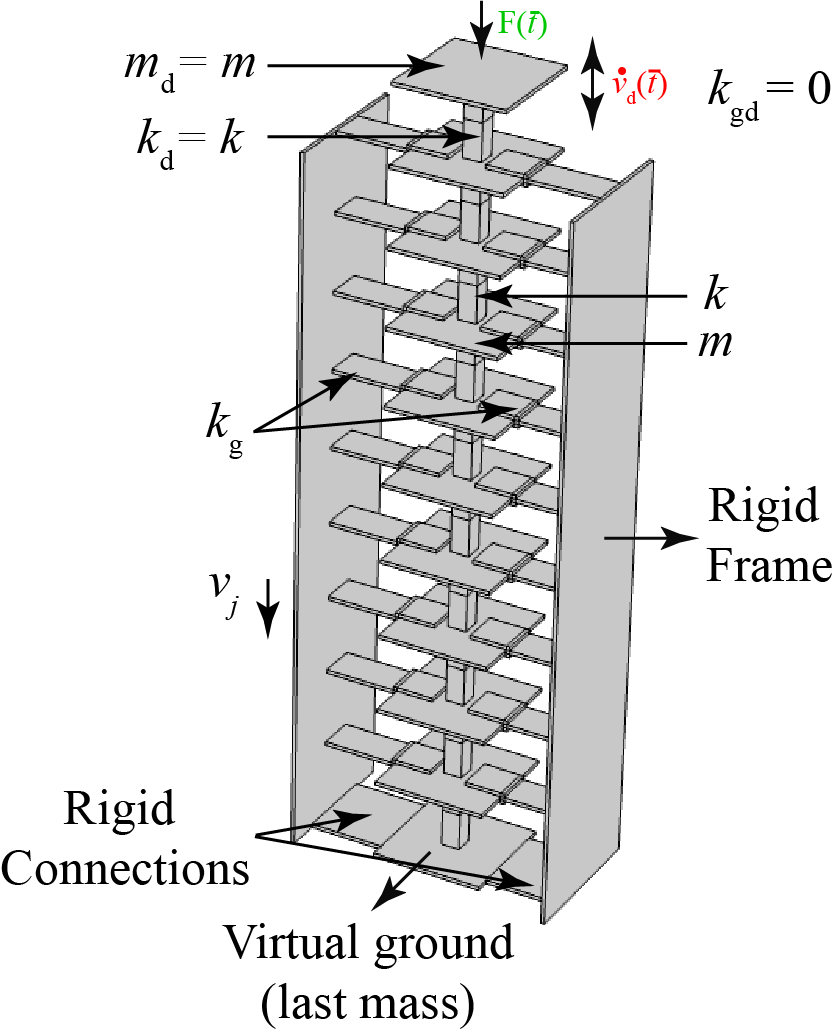}
    \caption{3D model of a PnC with a virtual ground.}
    \label{fig:SFIG4}
\end{figure*}

\section{PnC with a Virtual Ground}\label{sec:App_PnC_VG}
Fig. \ref{fig:SFIG4} shows a mechanical 3D model with rigid aluminum plates, elastic silicone beams, Veroclear grounding clips, and a rigid (metal) frame surrounding the periodic plate-beam-clip lattice, enabling a virtual grounding connection. The finite resonance frequencies of the PnCs with a virtual ground correspond to the eigenvalues of $\mathbf{K}_\mathrm{d}=\mathbf{K}_\mathrm{VG}-\Omega^2\mathbf{M}$, where the modified stiffness matrix, 
\begin{equation*}
    \mathbf{K}_\mathrm{VG}=\begin{bmatrix}
    k_\mathrm{r}+k_\mathrm{gd,r} & -k_\mathrm{r} & 0 & 0 & \dots & -k_\mathrm{gd,r}\\
    -k_\mathrm{r} & 1+k_\mathrm{r}+k_\mathrm{g,r} & -1 & 0 & \dots & -k_\mathrm{g,r}\\
    0 & -1 & 2+k_\mathrm{g,r} & -1 & \dots & -k_\mathrm{g,r}\\
    0 & 0 & -1 & \ddots & \ddots & \vdots\\
    \vdots & \vdots & \vdots & \ddots &  & -1-k_\mathrm{g,r}\\
    -k_\mathrm{gd,r} & -k_\mathrm{g,r} & -k_\mathrm{g,r} & \dots & -1-k_\mathrm{g,r} & 1+(\mathrm{N}-2)k_\mathrm{g,r}+k_\mathrm{gd,r}
    \end{bmatrix}_{\mathrm{N}\times\mathrm{N}},
\end{equation*}
and $\mathbf{M}=diag([m_\mathrm{r},1,1,\cdots,1,1+m_\mathrm{fr,r}]_{\mathrm{N}\times1})$. Connecting the rigid frame to the last mass adds an inertial mass, $m_\mathrm{fr}$ to the last mass. However, in Sec. \ref{sec:PnC_VG}, we assume a negligible frame mass (i.e., $m_\mathrm{fr}\approx0$), to illustrate the function of a virtual ground qualitatively. In the context of BG resonances, incorporating a finite grounding frame mass will primarily affect the characteristics of the high-frequency resonance at $\Omega=6.4095$, as the associated eigenmode mode is localized at the bottom boundary while leaving the edge mode ($\Omega_\mathrm{d,VG}=0.8447$) at the top boundary intact.

\section{FEM Frequency Analysis of the 3D defect-embedded structure}\label{sec:App_COMSOL}

\begin{figure*}
    \centering
    \includegraphics{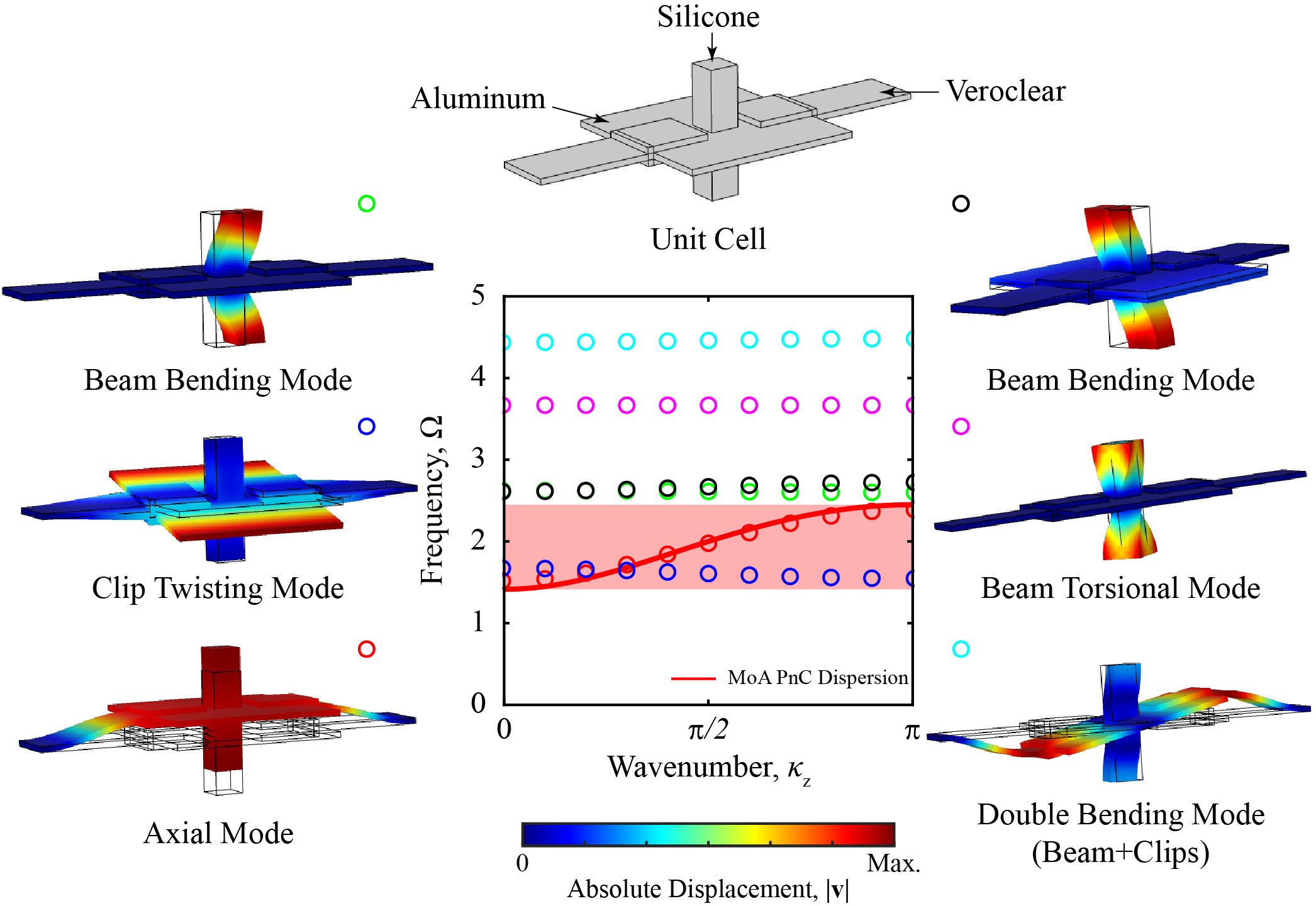}
    \caption{Dispersion relations of the 3D unit cell and the MoA PnC model. Different colored markers represent various branches of the dispersion curves of the 3D unit cell. Surrounding images depict the polarizations of the representative mode shapes of the dispersion branches. The solid red curve represents the dispersion relation derived from the MoA PnC model approximation for the axial dynamics of the 3D unit cell.}
    \label{fig:SFIG5}
\end{figure*}

\begin{table}[h]
    \centering
\begin{tabular}{|c|c|c|c|}\hline
     & Density, $\rho$ ($\mathrm{kg/m}^3$) & Young's Modulus, $E$ (GPa) & Poisson's Ratio, $\nu$ \\ \hline
     Aluminum & 2700 & 69 & 0.33\\ \hline
     Silicone & 1100 & 0.002 & 0.49\\ \hline
     Veroclear & 1180 & 3 & 0.4\\ \hline
\end{tabular}
     \caption{Material Properties of the 3D DMoA PnC model.}
     \label{tab:Table1}
\end{table}

A 3D model (Fig. \ref{fig:FIG1}(d-i)) consisting of aluminum plates ($m$), connected to nearest neighbors using silicone beams ($k$), and individually connected to a rigid ground using 3D Veroclear clips ($k_\mathrm{g}$) is chosen for the FEM study. The relevant material properties and the geometric dimensions of the constituent elements utilized in the study are given in Tab. \ref{tab:Table1} and Tab. \ref{tab:Table2}, respectively.

\begin{table}[h]
    \centering
\begin{tabular}{|c|c|c|c|}\hline
     & Length, $\ell$ (mm) & Breadth, $b$ (mm) & Height, $h$ (mm) \\ \hline
     Aluminum & 30 & 30 & 1\\ \hline
     Silicone & 5 & 5 & 20\\ \hline
     Veroclear (Cantilever portion) & 20 & 10 & 1\\ \hline
     \multicolumn{4}{|c|}{Lattice Parameter, $a = h_\mathrm{Al}+h_\mathrm{Sil} = 21 \ \mathrm{mm}$}\\ \hline
\end{tabular}
     \caption{Dimensions of the aluminum plate, silicone beam, and the Veroclear grounding clips.}
     \label{tab:Table2}
\end{table}

Before the 3D FEM frequency analysis, the axial dynamics are approximated as a lumped mass-spring reduced order model, and the inertial and elastic parameters are derived.

\begin{align}\label{eq:3D_Properties}
        m=\rho_\mathrm{Al}V_\mathrm{Al}&=\rho_\mathrm{Al}\ell_\mathrm{Al}b_\mathrm{Al}h_\mathrm{Al}=2700\times30\times30\times1\times 10^{-9}= 2.43 \times 10^{-3} \ \mathrm{kg},\\ \nonumber
        k&=\frac{E_\mathrm{Sil}A_\mathrm{Sil}}{h_\mathrm{Sil}}=\frac{E_\mathrm{Sil}\ell_\mathrm{Sil}b_\mathrm{Sil}}{h_\mathrm{Sil}}=\frac{2\times5\times5}{20}\times 10^{3}= 2500 \ \mathrm{N/m},\\ \nonumber
        k_\mathrm{g}=2\times\left(\frac{8E_\mathrm{Vero}I_\mathrm{Vero}}{\ell_\mathrm{Vero}^3}\right)&=2\times\left(\frac{8E_\mathrm{Vero}\left(\frac{b_\mathrm{Vero}\times h_\mathrm{Vero}^3}{12}\right)}{\ell_\mathrm{Vero}^3}\right)=2\times\frac{8\times3\times\left(\frac{10\times1^3}{12}\right)}{20^3}\times 10^{6}= 5000 \ \mathrm{N/m},
\end{align}
The stiffness of the grounding clips is assumed to be the bending stiffness of the cantilever portion, when subject to a uniformly distributed load over its surface\cite{TimoshenkoMOM1997}. Consequently, the reference temporal frequency, $\omega_0=\sqrt{k/m}=1014.3$ rad/s, and the non-dimensional grounding stiffness, $k_{g,r}=k_\mathrm{g}/k=2$. Fig. \ref{fig:SFIG5} contrasts the dispersion curves of the 3D FEM model and the lumped mass-spring MoA PnCs. The axial dispersion branch of the 3D model shows an excellent agreement with the MoA PnC dispersion curve.

The edge ($j=1$) and the bulk ($j=6$) defects are realized by omitting the grounding clips ($k_\mathrm{gd,r}=0$) at the respective defect locations in the 3D FEM model ($\mathrm{N}=11$). The Frequency Response Functions (FRFs) for the edge and bulk DMoA PnC configurations are obtained by applying a uniform boundary load to the defect lattice site (i.e., the aluminum plate) and monitoring the defect and boundary displacements.

\end{document}